\documentclass[10pt,journal,compsoc]{IEEEtran}

\ifCLASSOPTIONcompsoc
  \usepackage[nocompress]{cite}
\else
  \usepackage{cite}
  \fi
  
\usepackage[utf8]{inputenc} 
\usepackage{hyperref}       
\usepackage{url}            
\usepackage{booktabs}       
\usepackage{amsfonts}       
\usepackage{nicefrac}       
\usepackage{amsmath,amssymb,graphicx}
\usepackage{color}
\usepackage{mathrsfs,subfigure,epstopdf}
\usepackage{graphics,comment}
\usepackage{epsfig}

\makeatletter
\def\BState{\State\hskip-\ALG@thistlm}
\makeatother

\newcommand{\NN}{\mathbb{N}} 
\newcommand{\EE}{\mathbb{E}} 

\newcommand{\PP}{\mathbb{P}} 
\newcommand{\RR}{\mathbb{R}} 

\newcommand{\ee}{{\rm e}}
\newcommand{\dd}{{\rm\,d}} 

\newcommand{\qv}{{\bf q}}

\newcommand{\vvv}{{\bf v}}
\newcommand{\xv}{{\bf x}}

\newcommand{\zerov}{{\bf 0}}
\newcommand{\onev}{{\bf 1}}

\newcommand{\Am}{{\bf A}}

\newcommand{\Cm}{{\bf C}}
\newcommand{\Dm}{{\bf D}}

\newcommand{\Hm}{{\bf H}}
\newcommand{\Id}{{\bf I}}

\newcommand{\Mm}{{\bf M}}

\newcommand{\Sm}{{\bf S}}
\newcommand{\Tm}{{\bf T}}

\newcommand{\Zm}{{\bf Z}}

\newcommand{\Ac}{{\mathcal A}}
\newcommand{\Bc}{{\mathcal B}}

\newcommand{\Ec}{{\mathcal E}}

\newcommand{\Gc}{{\mathcal G}}

\newcommand{\Kc}{{\mathcal K}}

\newcommand{\Nc}{{\mathcal N}}

\newcommand{\Pc}{{\mathcal P}}
\newcommand{\Qc}{{\mathcal Q}}

\newcommand{\Sc}{{\mathcal S}}

\newcommand{\Wc}{{\mathcal W}}
\newcommand{\Vc}{{\mathcal V}}

\newcommand{\xiv}{\boldsymbol{\xi}}

\def\Tran{\mathsf{^T}}

\def\ben{\begin{enumerate}}
\def\beq{\begin{equation}}
\def\beqa{\begin{eqnarray}}
\def\bit{\begin{itemize}}
\def\een{\end{enumerate}}
\def\eeq{\end{equation}}
\def\eeqa{\end{eqnarray}}
\def\eit{\end{itemize}}

\def\non{\nonumber\\}
\def\argmax{\mathop{\mathrm{arg~max}}\limits}

\newcommand{\pf}{\noindent{\bf Proof:~}}
\newcommand{\qedsymb}{\hfill{\rule{2mm}{2mm}}}

\def\ben{\begin{enumerate}}
\def\beq{\begin{equation}}
\def\beqa{\begin{eqnarray}}
\def\bit{\begin{itemize}}
\def\een{\end{enumerate}}
\def\eeq{\end{equation}}
\def\eeqa{\end{eqnarray}}
\def\eit{\end{itemize}}

\def\non{\nonumber\\}
\def\argmax{\mathop{\mathrm{arg~max}}\limits}

\DeclareMathAlphabet{\mathsfbf}{OT1}{cmss}{sbc}{n}

\newtheorem{lemma}{Lemma}[section]

\newtheorem{definition}{Definition}[section]
\newtheorem{proposition}{Proposition}[section]
\newtheorem{example}{Example}[section]
\newtheorem{remark}{Remark}[section]

\begin{document}

\title{Ranking a set of objects: a graph based least-square approach}

 \author{
     Evgenia Christoforou,~\IEEEmembership{Member, IEEE},
      Alessandro Nordio,~\IEEEmembership{Member, IEEE},\\
      Alberto Tarable,~\IEEEmembership{Member, IEEE},
      Emilio Leonardi,~\IEEEmembership{Senior Member, IEEE}
  
\thanks{}
\IEEEcompsocitemizethanks{\IEEEcompsocthanksitem E. Cristoforou is with Research Centre on Interactive Media, Smart Systems and Emerging Technologies (RISE), Nicosia, Cyprus, email: evgenia.christoforou@gmail.com.%
  \IEEEcompsocthanksitem A.~Nordio and A. Tarable are with IEIIT-CNR (Institute of Electronics,
  Telecommunications and Information Engineering of the National Research Council of Italy),
  Italy, email: {firstname.lastname}@ieiit.cnr.it.%
  \IEEEcompsocthanksitem E. Leonardi is with DET,   Politecnico di Torino, Torino, Italy, email: emilio.leonardi@polito.it, and is associate researchers with IEIIT-CNR.}}%
  
\IEEEtitleabstractindextext{%
\begin{abstract}
  We consider the problem of ranking $N$ objects starting from a set
  of noisy pairwise comparisons provided by a crowd of equal workers.
  We assume that objects are endowed with intrinsic qualities and that
  the probability with which an object is preferred to another depends
  only on the difference between the qualities of the two competitors.
  We propose a class of non-adaptive ranking algorithms that rely on
  a least-squares optimization criterion for the
  estimation of qualities. Such algorithms are shown to be
  asymptotically optimal (i.e., they require
  $O(\frac{N}{\epsilon^2}\log \frac{N}{\delta})$ comparisons to be
  $(\epsilon, \delta)$-PAC).  Numerical results show that our schemes
  are very efficient also in many non-asymptotic scenarios exhibiting a
  performance similar to the maximum-likelihood algorithm. Moreover,
  we show how they can be extended to adaptive schemes and test them
  on real-world datasets.
\end{abstract}
}

\maketitle

\IEEEdisplaynontitleabstractindextext

%
\IEEEpeerreviewmaketitle

\section{Introduction}


Ranking algorithms have many applications. For example they are used
for ranking pages, users preferences against advertisements on the
web, hotels, restaurants, or online games~\cite{sponsoredads,
  NIPS2006_3079}.  In general a ranking algorithm infers an estimated
order relation among objects starting from a set of evaluations or
comparisons.  Sometimes, such evaluations are performed by human
``workers'' in the framework of crowdsourcing applications.  However,
since the behavior of humans cannot be deterministically predicted, it
is usually described through the adoption of a probabilistic
model. Then, the challenge in designing algorithms, is the ability to
infer reliable estimates of the ranking starting from ``noisy''
evaluations of the objects.  Often the ranking algorithm resorts to
pairwise comparisons of objects. In this work, we focus on such a
class of ranking algorithms.  Several stochastic models have been
proposed in the
literature~\cite{plackett1975analysis,luce2012individual,thurstone1927method,bradley1952rank}
to represent the outcome of comparisons.   Most of them are based on the idea that
objects to be compared have an intrinsic quality and that the
probability, $p_{i,j}$, that object $i$ is preferred to object $j$
depends on their qualities $q_i$ and $q_j$.  In this context, we
devise a class of efficient algorithms, which reconstruct object
qualities from pairwise difference through a least-square (LS) approach. To
do so, we establish a parallelism between the estimation process and
the average cumulative reward of random walks on a weighted graph.

\subsection{System model\label{sec:system_model}}
Let $\Qc\subset \RR$ be a compact set. We assume that $N$ objects are available for ranking: 
object $i$ is provided with an intrinsic quality, $q_i\in \Qc$, which is unknown to the
system.   Qualities induce a true ranking $r$ among objects, in which
$r(i) \prec r(j)$ iff $q_i>q_j$\footnote{The symbol $\prec$ is a
  precedence operator. If $r(i) \prec r(j)$ then object $i$
  ``precedes'' or ``is preferable to'' object $j$.}.  A ranking 
algorithm resorts to a set of observations (or answers) provided by
workers, which compare pairs of objects and return the identity of the
object they prefer. The comparison procedure implicitly contains some
randomness reflecting the workers' behavior. Thus, in general,
workers' answers can be modeled as a collection of binary random
variables, whose distribution depends on the qualities of the objects
to be evaluated.

Due to this randomness in the evaluation process, the inferred ranking
for object $i$, $\widehat{r}(i)$, does not always coincide with the
true ranking $r(i)$. The reliability of $\widehat{r}(i)$ depends on
how the evaluation process is organized. In particular, it depends on
(i) the workers' behavior, (ii) the choice of the set of object pairs
to be compared, (iii) the number of workers assigned to each pair of
objects, and (iv) the processing algorithm used to infer the ranking
from workers' answers.

We assume that all workers behave similarly and that they provide independent answers.
In particular, a worker comparing objects $i$ and $j$, will express a preference for
object $i$ against $j$ with probability:
\begin{equation}\label{eq:probability_model}
p_{i,j}= 1- p_{j,i}=F(q_i-q_j) 
\end{equation}
where the function $F(\cdot)$ is differentiable and 
strictly increasing in its argument (and therefore invertible) and such that
$F(0)=\frac{1}{2}$. Moreover, we assume that $F'(q)$ is bounded away from zero
for $q\in \bar{\Qc}$ where $\bar{\Qc} = \left\{ q_i-q_j | q_i, q_j\in \Qc \right\}$.
When the pair of objects $(i,j)$ is compared, the worker's output
is modeled as a binary random variable, $w_{i,j}\in[0,1]$, whose outcomes
have probability
\begin{equation}\label{eq:w}
\PP(w_{i,j}=1) = p_{i,j}; \qquad  \PP(w_{i,j}=0) = 1-p_{i,j}\,.
\end{equation}
The model in~\eqref{eq:probability_model} is pretty general. For
example, it encompasses
\begin{itemize}
\item the Thurstone model~\cite{thurstone1927method}, where the
  preferred object (in a pair) is chosen in accordance with the
  qualities {\em as perceived} by the worker and defined as
  \[ \widetilde{q}_i=q_i+n_i, \qquad \widetilde{q}_j=q_j+n_j \]
  respectively, where $n_i$ and $n_j$ are zero-mean random variables that represent
  noise terms. In this case $F(\cdot)$ is the cumulative distribution function
  of the zero-mean random variable $\eta_{i,j}=n_i-n_j$, i.e.,
  \begin{equation}\label{eq:p_Thurstone}
     p_{i,j} = \PP\left(\eta_{i,j} < q_i-q_j\right)
  \end{equation}

\item the  Bradley-Terry-Luce (BTL) model~\cite{plackett1975analysis,luce2012individual}, where
\begin{equation}\label{eq:p_Plackett-Luce}
  p_{i,j}=\frac{\ee^{q_i-q_j}}{1+\ee^{q_i-q_j}}\,.
\end{equation}

\end{itemize}
Let $\Vc = \{1,\dots,N\}$ be the set of objects. We observe that an
arbitrary choice of a set of object pairs to be compared, denoted by
$\Ec \subseteq \Vc \times\Vc$, automatically induces an undirected
graph $\Gc$, whose vertex and edge sets are, respectively,
$\Vc$ and $\Ec$. Clearly, it is possible to infer a ranking among the $N$ objects only if
the graph $\Gc$ is connected.

Each object pair $(i,j)\in \Ec$ is assigned to a number
of workers $W$.  In general, an increase of $W$ leads to a
more reliable estimate of the ranking. On the other hand, the overall
complexity, $C$, of the ranking algorithm is proportional to the 
total number of workers employed in the process, i.e., 
\[ C = |\Ec|W. \]
Then, an efficient ranking algorithm must find a
good trade-off between the complexity $C$ and the reliability of the
inferred ranking, i.e., by returning an almost correct ranking of
objects with a minimal number of pair comparisons.

About the reliability of the inferred ranking we say that an estimated
ranking is $\epsilon$-quality approximately correct (or, is an
$\epsilon$-quality ranking) if $\widehat r(i)\prec \widehat r(j)$
whenever $q_i\ge q_j - \epsilon$.  Moreover a ranking algorithm is
$(\epsilon,
\delta)$-PAC~\cite{szorenyi2015online,falahatgar2017maximum,falahatgar2018limits}
if it returns an $\epsilon$-quality ranking with
a probability larger than $1-\delta$.\footnote{Our definition of
	$(\epsilon, \delta)$-PAC algorithm slightly differs from the original given in 
	\cite{szorenyi2015online,falahatgar2017maximum,falahatgar2018limits} since it
	applies to object qualities. However, it can be easily shown to be asymptotically
	equivalent to the original. }

\subsection{Paper contribution and related work}
This paper contributes to a better
understanding of the fundamental limits of ranking algorithms based on
noisy pairwise comparisons.  Our main results complement and extend
previous findings about minimal complexity of ranking algorithms under
different non-parametric preference models recently derived in
\cite{,falahatgar2017maximum,falahatgar2018limits}.  As shown in
former studies the efficiency of ranking algorithm is crucially
determined by the structure of the underlying preference model.
 
On the one hand, under a non-parametric preference model satisfying
both Strong Stochastic Transitivity (SST) and Stochastic Triangle
Inequality (STI) properties,\footnote{A preference model is said non
  -parametric if pairwise preference probabilities are not
  necessarily induced by object qualities.  It satisfies the SST if
  $p_{i,k}\ge \max(p_{i,j},p_{j,k})$ whenever
  $r(i)\prec r(j)\prec r(k) $. It satisfies STI if
  $ p_{i,k}+\frac{1}{2}< p_{i,j}+ p_{j,k}$ whenever
  $r(i)\prec r(j)\prec r(k) $.} a provably asymptotically-optimal\footnote{A $(\epsilon, \delta)$-PAC ranking algorithm  is asymptotically-optimal  if its complexity is $O(\frac{N}{\epsilon^2}\log\frac{N}{\delta})$.}
{\em adaptive} algorithm has been proposed, under the restriction that
$\delta>\frac{1}{N}$. In particular, the algorithm proposed in
\cite{falahatgar2018limits} is $(\epsilon, \delta)$-PAC provided that
$O(\frac{N}{\epsilon^2}\log \frac{N}{\delta})$ comparisons are
dynamically allocated on the basis of previous outcomes. On the other
hand, in \cite{falahatgar2017maximum, falahatgar2018limits}, it
is shown that $\Omega(N^2)$ comparisons are strictly needed to obtain
a reliable ranking as soon as either STI or SST are relaxed.

When considering parametric models, estimating a ranking is essentially related to estimating the underlying qualities. \cite{shah2016estimation,
    graphresistance} provide a characterization of the
  expected norm-two distance between estimated and true qualities
  (later on referred to as mean square error (MSE)), in connection  with
  the properties of a fixed graph $\mathcal{G}$. In particular~\cite{shah2016estimation}, under the assumption that $F(\cdot)$ is
  log-concave, provides universal (i.e., applicable to optimal
  algorithms, such as the maximum-likelihood (ML) algorithm) order-optimal
  upper and lower bounds for the MSE, relating it to the
  spectral gap of a certain scaled version of the Laplacian of
  $\mathcal{G}$.  The very recent paper \cite{graphresistance}, for
  the BTL model only, introduces a LS algorithm and provides upper and lower
  bounds for a variant of the MSE and the relative
  tail-probabilities achievable by such algorithm, characterizing it in terms of the graph resistance. 
  
Interesting works are
  also~\cite{jang2016top,chen2015spectral,negahban2012iterative,negahban2016rank,DBLP:journals/corr/abs-1011-1716,DBLP:journals/corr/Cucuringu15,d2019ranking}.
  In
  ~\cite{DBLP:journals/corr/abs-1011-1716,DBLP:journals/corr/Cucuringu15}
  a LS approach for ranking is first introduced, but no theoretical
  guarantees are given.  In particular, \cite{DBLP:journals/corr/Cucuringu15} proposes Sync-Rank, a semi-definite programming algorithm based on the angular synchronization framework. In \cite{jang2016top,chen2015spectral},
  instead, an iterative algorithm that emulates a weighted random walk
  of graph $\Gc$ is proposed and its performance analyzed under the
  BTL model.  In particular, it provides bounds on the MSE and the
  corresponding tail-probabilities.  A direct comparison between the
  performance of algorithms proposed
  in~\cite{negahban2012iterative,negahban2016rank,graphresistance} is
  reported in ~\cite{graphresistance} where the LS approach is shown
  to be, in general, asymptotically more efficient.  Under
  the BTL model, ~\cite{jang2016top,chen2015spectral} propose and
  analyze algorithms able to identify the top-$k$ quality objects. At last,~\cite{d2019ranking}
  describes a ranking algorithm based on the singular value decomposition approach
  by assuming that workers return unquantized noisy estimates of objects
  quality differences.

Regarding online ranking algorithms, in \cite{szorenyi2015online}, for the BTL model, an online algorithm inspired to 
a finite-budget version of quick sort is described, able to obtain an  $(\epsilon, \delta)$-PAC ranking with 
$O(\frac{N}{\epsilon^2} \log N \log \frac{N}{\delta})$ comparisons. In \cite{ActiveRanking}, it is shown that, for online ranking algorithms, parametric models help to reduce the complexity only by logarithmic factors, in order sense. 

In this work, unlike \cite{graphresistance}, we introduce a rather general parametric preference model according to which preference probabilities are determined by an
arbitrary smooth monotonic function of object-quality differences. In this scenario, we
show that order-optimal {\em non-adaptive algorithms} can be defined
without the necessity of introducing any restriction to parameter
$\delta$.  In particular, differently from  \cite{shah2016estimation, graphresistance}, we work with the PAC framework and show that our algorithms are $(\epsilon, \delta)$-PAC,
provided that $O(\frac{N}{\epsilon^2} \log(\frac{N}{\delta} ))$
comparisons are blindly allocated in a single round.  Observe that our
preference model does not necessarily satisfy STI, while it satisfies
SST.  Our ranking procedure is based on the reconstruction of object
qualities from pairwise quality differences, by adopting a 
LS approach akin to the one in \cite{graphresistance}. Notice however that the analysis in \cite{graphresistance} only applies to the case where $\Omega(N \log^2(\frac{N}{\delta} ))$ total comparisons are performed. Our analysis establishes a parallelism between
the quality estimation process and the cumulative reward of random
walks on graphs. As an original contribution, we also introduce a \emph{weighted} LS algorithm with performance very close to the more complex ML algorithm. Finally, by simulation, we show that the performance
of our algorithms is extremely good also in
non-asymptotic scenarios. 

  
    The paper is organized as follows: in Section~\ref{subsec:ML} we
introduce a ranking algorithm based on the Maximum Likelihood (ML) approach, which is used
as a performance reference.  In Section~\ref{subsec:LS}
we describe our proposed LS estimation algorithm, whose
asymptotic analysis is investigated in
Section~\ref{sec:asymptotic_analysis}.  The LS estimation algorithm is
then tested in Sections~\ref{sec:res_synthetic}
and~\ref{sec:results_real_world} against synthetic and real-world
datasets, respectively. Finally, in Section~\ref{sec:conclusions} we
draw our conclusions.

\subsection{Notation}
Boldface uppercase and lowercase letters denote matrices and vectors,
respectively. $\Id$ is the identity matrix. The transpose of the
matrix $\Am$ is denoted by $\Am\Tran$, while $[\Am]_{i,j}$ indicates its
$(i,j)$-th entry. For the sake of notation compactness we 
use the notation $\Am=\{a_{i,j}\}$ to define a matrix $\Am$ whose elements are $a_{i,j}$.
Finally, the symbol $\odot$ represents the Hadamard product, while calligraphic letters denote sets or graphs.

\section{Maximum-likelihood quality estimation\label{subsec:ML}}
Consider a graph $\Gc(\Vc,\Ec)$ with $|\Vc|=N$ vertices where each
pair of objects $(i,j)\in\Ec$ is evaluated $W$ times by independent
workers~\footnote{The generalization to a number of
  evaluations that depends on the specific edge is straightforward.}.
Without loss of generality
we assume that the indices of the objects connected by the generic
edge $(i,j)\in\Ec$ are such that $i>j$. Moreover, we assume that the
$m$-th worker evaluating the pair of objects $(i,j)$ outputs the
binary random variable $w_{i,j,m}$ whose distribution is given
by~\eqref{eq:w}.

In our proposed ML approach, the estimate of the ranking can be
obtained by sorting the quality estimates
$\widehat{\qv}=[\widehat{q}_1,\ldots,\widehat{q}_N]\Tran$ which are obtained as follows:
\begin{equation}
  \widehat{\qv}
   \mathord{=} \argmax_{\qv} \log \PP\left(\{w_{i,j,m}, (i,j)\in \Ec, m=1,\ldots,W\}| \qv\right)\,. \label{eq:ML}
\end{equation}
When workers are independent on each other and behave similarly,
the random variables $w_{i,j,m}$ can be modeled as independent and
identically distributed. Therefore, the conditional probability in~\eqref{eq:ML} factorizes as
\begin{eqnarray}
 && \hspace{-18ex}\PP\left(\{w_{i,j,m}, (i,j)\in \Ec, m=1,\ldots,W\}| \qv\right) \non
\hspace{+6ex}  &\hspace{+6ex}=&\prod_{(i,j)\in \Ec}\prod_{m=1}^{W}\PP\left(w_{i,j,m}| \qv\right).\nonumber \label{eq:ML_factorization}  
\end{eqnarray}
By using~\eqref{eq:w} we write
\[ \PP\left(w_{i,j,m}| \qv\right) = p_{i,j}^{w_{i,j,m}}\left(1-p_{i,j}\right)^{1-w_{i,j,m}}\,.\]
where we recall that $p_{i,j}=F(q_i-q_j)$.
By substituting the above result in~\eqref{eq:ML}, the ML estimate of
the qualities $\qv$ can be rewritten as
\begin{eqnarray}
  \widehat{\qv}&\hspace{-2ex}=&\hspace{-2ex} \argmax_{\qv} \log \PP\left(\{w_{i,j,m}, (i,j)\in \Ec, m=1,\ldots,W\}| \qv\right) \non
  &\hspace{-2ex}=&\hspace{-2ex} \argmax_{\qv} \sum_{(i,j)\in \Ec}\sum_{m=1}^{W}\log p_{i,j}^{w_{i,j,m}}\mathord{+}\log\left[(1\mathord{-}p_{i,j})^{(1\mathord{-}w_{i,j,m})}\right]\non
  &\hspace{-2ex}=&\hspace{-2ex} \argmax_{\qv}\Psi(\qv)
\end{eqnarray}
where
\[\Psi(\qv)=\sum_{(i,j)\in \Ec}s_{i,j}\log
  p_{i,j}+(1-s_{i,j})\log(1-p_{i,j})\,.\]
and $s_{i,j}=\frac{1}{W}\sum_{m=1}^Ww_{i,j,m}$. The function $\Psi(\qv)$ has a finite global maximum. Indeed, since
$p_{i,j}\in[0,1]$, and $s_{i,j}\in[0,1]$, it is straightforward
to show that $\Psi(\qv)\le 0$. However, in general, $\Psi(\qv)$ is a
non-linear function of $\qv$ and its maximization non trivial.
Nevertheless, a local maximum can be found by using standard techniques such
as, for example, the Newton-Raphson method which works
iteratively and requires the function $F(\cdot)$ to be twice differentiable.

Let $\widehat{\qv}_t$ be the estimate of $\qv$ at iteration
$t=1,2,\ldots$.  Then the estimate of $\qv$ at iteration $t+1$ can be updated
as follows:
\[ \widehat{\qv}_{t+1} = \widehat{\qv}_{t} - [\Sm(\widehat{\qv}_{t})]^{-1}\nabla
  \Psi(\qv)_{t}\]
where $\nabla \Psi(\qv)$ and $\Sm(\qv)$ are, respectively, the gradient and the Hessian matrix of $\Psi(\qv)$.
Specifically,  $[\nabla\Psi(\qv)]_h \triangleq  \frac{\partial \Psi(\qv)}{\partial q_h}$ and $[\Sm(\qv)]_{h,k} \triangleq \frac{\partial^2 \Psi(\qv)}{\partial q_h\partial q_k}$.
In order to compute $\nabla \Psi(\qv)$ and $\Sm(\qv)$
consider a generic node $h\in \Vc$ and the set $\Ec_h\subseteq \Ec$ of edges connecting node $h$ to its neighbors.
Then, the function $\Psi(\qv)$ can be rewritten as
\begin{equation}
  \Psi(\qv)  = c+ \sum_{(i,j)\in \Ec_h}s_{i,j}\log p_{i,j}+(1-s_{i,j})\log(1-p_{i,j})
\end{equation}
where the term $c$ does not depend on $q_h$. Since $p_{i,j}=F(q_i-q_j)$, we can write the partial derivatives of $p_{i,j}$ as follows:
\[ \frac{\partial p_{i,j}}{\partial q_i} \triangleq p'_{i,j}; \quad \frac{\partial p_{i,j}}{\partial q_j} \triangleq -p'_{i,j}\]
and, similarly
\[ \frac{\partial^2 p_{i,j}}{\partial q_i^2}=\frac{\partial^2 p_{i,j}}{\partial q_j^2}\triangleq p''_{i,j}; \qquad \frac{\partial^2 p_{i,j}}{\partial q_j\partial q_j}= -p''_{i,j}.\]
It immediately follows that
\begin{eqnarray}
 [\nabla\Psi(\qv)]_h 
 &=&  \sum_{(h,j)\in \Ec_h}p'_{h,j}\left[\frac{s_{h,j}-p_{h,j}}{p_{h,j}(1-p_{h,j})}\right]\non
      &&\quad -\sum_{(i,h)\in \Ec_h}p'_{i,h}\left[\frac{s_{i,h}-p_{i,h}}{p_{i,h}(1-p_{i,h})}\right]\nonumber
\end{eqnarray}
and
\begin{eqnarray}
  [\Sm(\qv)]_{h,h}
  &=&  \sum_{(h,j)\in \Ec_h}p''_{h,j}\left[\frac{s_{h,j}-p_{h,j}}{p_{h,j}(1-p_{h,j})}\right]\non
  &&\quad-(p'_{h,j})^2\left[\frac{p_{h,j}^2+s_{h,j}(1-2p_{h,j})}{p_{h,j}^2(1-p_{h,j})^2}\right]\non
  &&\quad+\sum_{(i,h)\in \Ec_h}p''_{i,h}\left[\frac{s_{i,h}-p_{i,h}}{p_{i,h}(1-p_{i,h})}\right]\non
  &&\quad-(p'_{i,h})^2\left[\frac{p_{i,h}^2+s_{i,h}(1-2p_{i,h})}{p_{i,h}^2(1-p_{i,h})^2}\right].\nonumber    
\end{eqnarray}
Moreover, for $h\neq k$
\begin{eqnarray}
  [\Sm(\qv)]_{h,k} 
  \mathord{=}\left\{ \begin{array}{l}
            0 \\ \qquad\qquad \mbox{if } (h,k)\notin \Ec_h \mbox{ or } (k,h)\notin \Ec_h \\
            (p'_{h,k})^2\frac{p_{h,k}^2+s_{h,k}(1-2p_{h,k})}{p_{h,k}^2(1-p_{h,k})^2}\mathord{-}p''_{h,k}\frac{s_{h,k}-p_{h,k}}{p_{h,k}(1-p_{h,k})} \\
            \qquad\qquad \mbox{if } (h,k)\in \Ec_h \mbox{ or } (k,h)\in \Ec_h
          \end{array}\right.\nonumber                                                                                                                        
\end{eqnarray}

The above equations can be specialized for both the Thurstone model as
well as for the BTL model, by using the expressions for $p_{i,j}$
provided, respectively, in~\eqref{eq:p_Thurstone} and~\eqref{eq:p_Plackett-Luce}.

\section{Least-squares quality estimation\label{subsec:LS}}
We propose a simpler linear estimation algorithm, based on a least-square criterion, that can be applied on the graph $\Gc(\Vc,\Ec)$.
Let the distance between objects $i$ and $j$ be
\[ d_{i,j} = q_{i}-q_{j} \]
and let $\Wc_{i,j}$ be the set of binary answers, of cardinality $W$, provided by the workers comparing the pair $(i,j)$.
Also, let $K_{i,j}$ be the number of times object $i$ is preferred to
object $j$. Then, by construction, $K_{i,j}$ follows the binomial distribution $K_{i,j} \sim \mathrm{Bin}(W, p_{i,j})$,
where $p_{i,j} = F(d_{i,j})$. Out of the evaluation results,
an estimate $\widehat{d}_{i,j}$ of $d_{i,j}$ is formed as
\begin{equation} \label{eq:d_hat}
  \widehat{d}_{i,j}= F^{-1}\left(\widehat{p}_{i,j}\right)=F^{-1}(y_{i,j}+p_{i,j})\,,
\end{equation}
where $\widehat{p}_{i,j}=K_{i,j}/W$ is the estimate of
$p_{i,j}$, and $y_{i,j}=\widehat{p}_{i,j}-p_{i,j}$ represents the
estimation error on the probability $p_{i,j}$.  Note that $y_{i,j}$
has zero mean and variance $\EE[y^2_{i,j}]= \frac{p_{i,j}(1-p_{i,j})}{W}$.
As a consequence, $\widehat{d}_{i,j} = d_{i,j}+ z_{i,j}$, where $z_{i,j}$ represents the error on the estimate of $d_{i,j}$
induced by the presence of $y_{i,j}$.
From the set of noisy estimates $\{\widehat{d}_{i,j}, (i,j)\in \Ec\}$, the estimate $\widehat{\qv}=[\widehat{q}_1,\ldots,\widehat{q}_N]\Tran$ of $\qv=[q_1,\ldots,q_N]\Tran$ can be obtained by solving the following LS
optimization problem
	\begin{equation}\label{min-prob}
\widehat{\qv} = \arg \min_{\xv}  \sum_{(i,j) \in \Ec} \omega_{i,j}\left(x_i-x_j - \widehat{d}_{i,j} \right)^2 
\end{equation}
where $ \omega_{i,j}$ are arbitrary positive weights,  whose setting  is discussed in Section~\ref{sec:weights}.
The  solution of \eqref{min-prob} satisfies the following linear equations:
  \begin{equation}\label{syst-eq}
  \widehat{q}_i  = \sum_{j\in \mathcal{N}_i} \omega_{i,j} \frac{\widehat{q}_j+\widehat{d}_{i,j}}{\rho_i}, \,\,\, i=1,\dots,N
\end{equation}
where $\mathcal{N}_{i}$ represents the neighborhood of node $i$ (i.e.,
the set of nodes connected to $i$ in $\Gc$), and $\rho_i$ is
its  generalized degree, i.e., $\rho_i =\sum_{j\in \mathcal{N}_i}\omega_{i,j}$.  We can compactly
express the previous linear system in terms of the $N \times N$
 matrix $\widetilde{\Hm}$ associated to the graph
$\Gc$, whose elements are defined as
\[
[\widetilde{\Hm}]_{i,j} = \left\{ \begin{array}{cc} 
\omega_{i,j}/{ \rho}_{i}, & (i,j) \in \Ec, \\
0, & \mbox{otherwise.}
\end{array}
\right.
\]
Let $\Id$ be the identity matrix,
$\widetilde{\Mm} = \Id - \widetilde{\Hm}$, and $\Zm =
\{z_{i,j}\}$. Moreover let $\Dm = \{ d_{i,j }\}$ and
$\widehat{\Dm}= \{ \widehat{d}_{i,j} \}$ be, respectively, the
antisymmetric matrices of the true and estimated quality
differences\footnote{Notice that the $(i,j)$-th entry of $\widehat{\Dm}$
  is defined only for $(i,j) \in \Ec$. The same is true for matrix
  $\Zm$.}.  Thus, from~\eqref{syst-eq} we can write:
\begin{equation}\label{eq:linear_system}
\widetilde{\Mm} \widehat{\qv} = (\widetilde{ \Hm} \odot \widehat{\Dm}) \onev =  (\widetilde{ \Hm} \odot \Dm) \onev+(\widetilde{\Hm} \odot \Zm )\onev\,.
\end{equation}
where $\odot$ represents the Hadamard product and
$\onev=[1,\ldots,1]\Tran$ is a column vector of size $N$.  We observe
that, by construction, $\mathrm{rank}(\widetilde{\Mm}) = N-1$, i.e.,
$\widetilde{\Mm}$ is singular. Indeed $\widetilde{\Mm} \onev=\zerov$,
as it can be easily checked. This implies that the associated linear
operator on $\RR^N$ is not injective and that, given a solution
${\widehat{\qv}'}$ of \eqref{syst-eq}, also
$\widehat{\qv}'' = \widehat{\qv}' +\alpha \onev$ is a solution
of~\eqref{syst-eq} for any $\alpha\in \RR$. Note, however,
that, for the purposes of object ranking, the actual value of $\alpha$
is irrelevant, since every solution of the form
$\widehat{\qv}'' = \widehat{\qv}' +\alpha \onev$ induces the same
object ranking.  Therefore, we can arbitrarily fix the quality of,
say, object $N$ to 0 as a reference, i.e., $q_N = 0$.  To keep into
account this constraint, we define the new matrices $\Hm$ and $\Mm$ as
follows:
\begin{equation} [\Hm]_{i,j}= \left\{
\begin{array}{ll}
  [\widetilde{\Hm}]_{i,j} & i<N,\,\, \forall j \\
  0                      & i=N,\,\,\forall j 
\end{array} 
\right. 
\end{equation}
and $\Mm = \Id-\Hm$, respectively. We then replace $\widetilde{\Mm}$
and $\widetilde{\Hm}$ in~\eqref{eq:linear_system} with, respectively,
$\Mm$ and $\Hm$. Since $\Mm$ is full rank, solving for $\qv$ we obtain
\begin{equation} \label{eq:final}
\widehat{\qv}  = \Mm^{-1} (\Hm \odot \widehat{\Dm}) \onev = \qv +  \Mm^{-1} (\Hm \odot \Zm) \onev 
\end{equation} 	
where we have used the fact that $\qv =  \Mm^{-1} ({\Hm } \odot \Dm) \onev$.

\subsection{Weight optimization\label{sec:weights}}

In the following, we will consider two possible choices for the weights $\omega_{i,j}$. The first, which will be studied in the next section for its simplicity, corresponds to $\omega_{i,j} = 1$ for all $i,j$, and will be called unweighted LS or simply LS. The second, which will be called weighted LS (WLS), is dictated by the fact that the estimates $ \widehat{d}_{i,j}$ do not have the same reliability. Indeed, by developing 
\eqref{eq:d_hat} at the first order for $W \rightarrow \infty$, we obtain
\[z_{i,j} = \widehat{d}_{i,j} - d_{i,j} = \left.\frac{d F^{-1}(p)} {d p}\right|_{p=p_{i,j}} y_{i,j}+O\left(y_{i,j}^2 \right)\,.\]
so that, if we neglect the higher-order term, $z_{i,j}$ is a zero-mean random variable with variance 
\[
\sigma^2_{i,j} = \left(\left.\frac{d F^{-1}(p)} {d p}\right|_{p=p_{i,j}} \right)^2  
\frac{p_{i,j}(1-p_{i,j})}{W} \]
Given the values of $q_j$, $j \in \Nc_i$, the optimal weights for $W \rightarrow \infty$ in \eqref{syst-eq} are then proportional to $\sigma^{-2}_{i,j}$. For our WLS algorithm, we will then set $\omega_{i,j} = \widehat{\sigma}^{-2}_{i,j}$ , with
 \[
\widehat{\sigma}^2_{i,j} = \left(\left.\frac{d F^{-1}(p)} {d p}\right|_{p=\widetilde{p}_{i,j}} \right)^2 . 
\frac{\widetilde{p}_{i,j}(1-\widetilde{p}_{i,j})}{W}, \]
where $\widetilde{p}_{i,j}=\max (\min(\widehat{p}_{i,j},1-\chi), \chi)$,  for a small positive parameter $\chi$ such that $\left. \frac{d F^{1}(p)} {d p}\right|_{p=\chi}$ exists finite.
Note that, under this setting $0<\omega_{i,j}<\infty.$
\section{Asymptotic analysis of the Least-Square Estimator\label{sec:asymptotic_analysis}}

All the theoretical results in this section are obtained by
considering the unweighted LS estimator, for simplicity. However,
they can be extended to the general weighted
case as long as $\min_{ij} \omega_{i,j}/\max_{ij} \omega_{i,j}$ is
bounded away from 0, as for the case described in Section~\ref{sec:weights}.

The following propositions derive the conditions for the asymptotic
convergence of the estimated qualities to their true values. 
We start by presenting  a preliminary  asymptotic result  on the  	mean square error.	
\begin{proposition} \label{second-mom} Consider the unweighted LS estimator in \eqref{syst-eq}. Assume that the
  degree of nodes of the graph are upper-bounded and define $\rho_{\inf} \triangleq \inf_i \rho_i$.
  Then the mean square error (MSE) on the estimates $\hat{\qv}$ can be bounded by
  \begin{equation}\label{eq:secmom}
  \EE[(\widehat{\qv}-\qv)\Tran (\widehat{\qv}-\qv) ] \leq c\lambda_\Cm^{\rm max}\frac{N}{W\rho_{\inf}}
  \end{equation}
  where $c$ is a constant, 
  $\lambda^{\max}_{\Cm}$ is the largest eigenvalue of
  $\Cm=\left(\Mm^{-1}\right)\Tran\Mm^{-1}$ and $W \ge \beta \log N$ for a
  sufficiently large $\beta$.
\end{proposition}
The proof is provided in Appendix~\ref{app:second-mom}.
\medskip


Even if an expression similar to~\eqref{eq:secmom} is reported in
\cite{shah2016estimation}, we recall that the latter was derived for
perfect ML-estimators under the assumption that $F(\cdot)$ is
log-concave; our results, instead, apply to LS algorithm for a generic
strictly-increasing $F(\cdot)$. Furthermore, \eqref{eq:secmom}
complements and extends results in \cite{graphresistance} under more
general settings (we recall that results in \cite{graphresistance}
apply to the BTL model only). It is also to be noted that the theoretical results in \cite{graphresistance} only apply to the regime where $W$ is large, i.e., $W = \Omega \left(\log^2 \frac{N}{\delta} \right)$. Under such constraint, for any connected graph, the total complexity of the algorithm in terms of number of comparisons is at least $\Omega \left(N \log^2 \frac{N}{\delta} \right)$.

From~\eqref{eq:secmom}, we can deduce that, whenever
$\lambda_{\Cm}^{\max}$ is bounded (as for example in the case of
Ramanujan graphs), by symmetry,
$\EE[(\widehat{q}_i- q_i)^2]= O\left(\frac{1}{W\rho_{\inf}}
\right)$, $i = 1,\dots,N$. Thus, if
$W \rightarrow \infty$ for $N\rightarrow \infty$, then
$\widehat{\qv}$ converges in probability to $\qv$.  

 To find out the
minimum number of comparisons under which the LS approach satisfies
the $(\epsilon, \delta)$-PAC conditions, we need to evaluate
$\PP(\sup_i |\widehat{q}_i- q_i| > \epsilon)$ for $\epsilon>0$.  The
following proposition gives sufficient conditions in order for the
absolute error to converge to zero in the properly defined limiting
regime.
\begin{proposition} \label{EPAC} Consider the unweighted LS estimator in \eqref{syst-eq}. For any $\epsilon>0$, as
  $N$  grows, $\PP(\sup_i |\widehat{q}_i- q_i| > \epsilon)<\delta$,
  provided that
\begin{enumerate}
\item[i)] $\limsup_{N\to \infty}\|\Mm^{-1}\|_\infty<\infty$ (i.e., the $\infty$-norm  of $\Mm^{-1}$ is bounded),
\item[ii)] the  total number of edges of $\Gc$ is $O(N)$, and $W >\beta(\epsilon,\delta) \log N $ for some 
$\beta(\epsilon,\delta)=O\left(\frac{1}{\epsilon^2}\frac{\log\frac{N}{\delta}}{\log N}\right)$.
\end{enumerate}
Assumption i) can be weakened by the following condition i'):
\begin{enumerate}
\item[i')] $\limsup_{N\to \infty} \sup_{\Am: \|\Am\|_{\infty}\le 1 } \| \Mm^{-1}(\Hm \odot \Am ) \onev\|_\infty <\infty$.
\end{enumerate}
\end{proposition}
The proof is provided in Appendix~\ref{app:EPAC}.
\medskip

\begin{remark}
	Note that Proposition \ref{EPAC} provides sufficient conditions for the existence of a $(\epsilon,\delta)$-PAC ranking algorithm with 
	complexity $O(\frac{N}{\epsilon^2}\log\frac{N}{\delta})$.  In the following subsection we characterize classes of graphs   meeting  condition (i) or  (i') of Proposition~\ref{EPAC}.
\end{remark}

\subsection{Considerations on graphs structure}
Proposition~\ref{EPAC} grants that the absolute error
$\sup_i|\widehat{q}_i-q_i|$ can be well controlled as $N\to \infty$
under some conditions on the matrix $\Mm$ (condition (i) or (i')).
Such conditions hold depending on the structure of the graph $\Gc$.
In order to characterize the class of graphs for which condition (i)
or condition (i') holds, we first observe that~\eqref{eq:final}
computes the quality of object $i$ as the average value of the sum of
estimated quality differences along all paths joining node $i$ to the
reference node $N$. In other words, $\widehat{q}_i$ can be regarded as
the average total reward earned by a standard random walk that starts
from node $i$ and stops as soon it hits node $N$, when estimates
$\widehat{d}_{i,j}$ are the elementary rewards associated to graph
edges \cite{ewaldMC}. Then, in Section~\ref{sec:direct} we restrict our
asymptotic analysis to directed and acyclic graphs. Finally, in
Section~\ref{sec:undirected} we extend it to the more general class of
undirected graphs.

\subsubsection{$\Gc$ is directed and acyclic\label{sec:direct}}
An explicit solution of~\eqref{syst-eq} can be given when graph
$\Gc$ is turned into a directed graph, i.e., by imposing to all edges
one of the two possible directions.  While this assumption is
suboptimal, since it constrains the random walk to a subset of possible trajectories, it greatly
simplifies the analysis. Indeed, first observe that, for directed
graphs, \eqref{syst-eq} can be rewritten as:
\[
\widehat{q}_i  = \sum_{j\in \Nc^-_i}\frac{\widehat{q}_j+ \widehat{d}_{i,j}}{\rho^-_i}.
\] 
where $\Nc^-_i$ represents the set of in-neighborhoods of $i$ and
$\rho^-_i = |\Nc^-_i|$.  Then, when the graph is directed and
acyclic, and has the reference node, $N$, as a common ancestor, an
explicit solution for $\widehat{q}_i$, $i = 1,\dots,N-1$, is
\begin{eqnarray}\label{explicit}
  \widehat{q}_i= \frac{1}{\rho^-_i}  \sum_{j_1\in\Nc^-_i} \left[\widehat d_{i,j_1}+ \frac{1}{\rho^-_{j_1}}\sum_{j_2\in\Nc^-_{j_1}}\gamma_{j_2} \right]
\end{eqnarray}
where
\[\gamma_{j_2} = \widehat d_{j_1,j_2}+\ldots +\frac{1}{\rho^-_{j_{\ell_{i}-1}}}\sum_{j_{\ell_i}\in\Nc^-_{j_{\ell_{i}-1}}}\widehat{d}_{j_{\ell_{i}-1} j_{\ell_{i}}} \]
and $\ell_i$ is the length of the longest (simple) path from node $1$ to
the reference node.
Proposition~\ref{prop:normgraphsdir} gives sufficient conditions for a directed
acyclic graph to meet the requirements of Proposition \ref{EPAC}. The proposition exploits the notion of proximality
between nodes according to the following definition:
\begin{definition}
  Given a family of graphs $\{\Gc_N\}_N$, we say that a node $i$ is
  {\em proximal} to the reference node $N$, with parameters $(\tau,h)$, if a random walk starting
  from $i$ reaches the reference node $N$ within $h$ hops with a
  probability that is asymptotically (with $N$) bounded below by $\tau$.
\end{definition}

\begin{proposition}\label{prop:normgraphsdir}
  Given a family of directed and acyclic graphs
  $\{\widehat{\Gc}_N\}_N$ with bounded diameter, condition (i') of
  Proposition \ref{EPAC} is satisfied if one of the following three
  conditions is met: (i) all paths from any node to the reference have
  bounded length, (ii) $\sup_i \rho_i^{-} < \infty$, or (iii) a
  fraction bounded away from 0 of the in-neighbors of any node is
$(\tau,h)$ proximal for some $\tau>0$ and $ h<\infty$.
\end{proposition}
The proof is provided in Appendix~\ref{app:normgraphsdir}.
\medskip

\subsubsection{$\Gc$ is undirected\label{sec:undirected}}
Now, let us go back to the original formulation \eqref{syst-eq} on the
undirected graph. In the following, we will show that, considered from
the point of view of a given node, the solution of \eqref{syst-eq} for an undirected graph
can be obtained by defining an equivalent problem for a properly defined
directed acyclic graph. Consider the graph $\Gc = (\Vc, \Ec)$ on $N$
nodes and let $\Tm$ be the $(N-1) \times (N-1)$ matrix obtained from
matrix $\Hm$ by removing the last row and column (i.e., those
corresponding to the reference node $N$). Consider a given node $i$,
$i=1,\dots, N-1$, and notice that $\left[(\Id- \Tm)^{-1}\right]_{i,j}$
gives the average number of times that node $j$ is visited in the
random walk starting from $i$, before ending in the reference node
$N$ \cite{kemeny1976markov}. Let
\begin{equation}\label{eq:theta_def}
\theta_{j,i} = \frac{\left[(\Id- \Tm)^{-1}\right]_{i,j}}{\rho_j}
\end{equation} 
be the average number of times any edge incident to node $j$ is
traversed in the direction from $j$ to its neighbors, in the standard
random walk defined  on $\Gc$.  Now, define a directed graph
$\stackrel{\rightarrow}{\Gc}_i = (\Vc,
\stackrel{\rightarrow}{\Ec}_i)$, where
$(j,\ell) \in \stackrel{\rightarrow}{\Ec}_i$ if and only if
$(j,\ell) \in \Ec$ and one of the two following conditions are
satisfied: (i) $j < N$ and $\ell = N$, or (ii) $j < N$, $\ell < N$,
and $\theta_{j,i} > \theta_{\ell,i} $.

Notice that $\stackrel{\rightarrow}{\Ec}_i \subset \Ec$.  Let
$\Nc_{ji}^{-}$ be the set of in-neighbors of $j$ in
$\stackrel{\rightarrow}{\Gc}_i$. It can be easily verified that in
$\stackrel{\rightarrow}{\Gc}_i$ node $i$ has only in-neighbors and
node $N$ (the reference node) has only out-neighbors. It is also easy
to prove that $\stackrel{\rightarrow}{\Gc}_i$ is acyclic. Indeed,
suppose that the cycle $(j_1, j_2, \dots, j_r, j_1)$ belongs to
$\stackrel{\rightarrow}{\Gc}_i$. This implies that, by definition,
$\theta_{j_1,i} > \theta_{j_2,i} > \dots > \theta_{j_r,i} > \theta_{j_1,i}$, which is impossible.

Let us also define a biased random walk on digraph $\stackrel{\rightarrow}{\Gc}_i$,
for which, given that the current node is $j$, the probability of
taking outgoing edge $(j, \ell)$, $\ell \in \Nc_{ji}^{-}$ is given by
\begin{equation}\label{eq:bias_prob}
  \eta_{j \rightarrow \ell, i} = \frac{\theta_{j,i}
  - \theta_{j,\ell}}{\sum_{\ell' \in \Nc_{ji}^{-}} \left( \theta_{j,i} -
    \theta_{j,\ell'}\right)}
\end{equation}
 The following proposition relates
the standard random walk on $\Gc$ to the biased random walk on
$\stackrel{\rightarrow}{\Gc}_i $.

\begin{proposition} \label{prop:equivdirundir} The estimate of $q_i$
  given by \eqref{eq:final} on $\Gc$ can be obtained by solving
\begin{equation} \label{eq:syst_eq_dir}
\check{q}_j = \sum_{\ell \in \Nc_{ji}^-} (\check{q}_{\ell}+\widehat{d}_{j,\ell})\eta_{j \rightarrow \ell, i}, \,\,\, i=1,\dots,N
\end{equation}
on $\stackrel{\rightarrow}{\Gc}_i$, and then setting $\widehat{q}_i = \check{q}_i$.
\end{proposition}
The proof is provided in Appendix~\ref{app-prop:equivdirundir}.
\medskip

According to Proposition~\ref{prop:equivdirundir}, $\widehat{q}_i$ can be equivalently seen as the
average total reward of the standard random walk on graph $\Gc$ or as
the average total reward of the biased random walk on graph
$\stackrel{\rightarrow}{\Gc}_i$. The following proposition gives
sufficient conditions for a family of graphs to meet the conditions of
Proposition \ref{EPAC}.
  
\begin{proposition}\label{prop:normgraphsundir}
  Given a family of graphs $\{\Gc_N\}_{N \in \NN}$ with bounded
  diameter, condition i') of Proposition \ref{EPAC} is satisfied if,
  for each node $i$, $i = 1,\dots,N-1$ one of the following conditions
  are satisfied: (i) all paths in $\stackrel{\rightarrow}{\Gc}_i$ from
  $i$ to the reference have bounded length, (ii) in
  $\stackrel{\rightarrow}{\Gc}_i$, a fraction bounded away from 0 of
  the in-neighbors of any node is proximal.
\end{proposition}
The proof is provided in Appendix~\ref{app-prop:normgraphsundir}.
\medskip

\begin{example}
  Consider the family of complete graphs on $N$ nodes, i.e.,
  $\Gc_N = \Kc_N$.\footnote{Note that even if  this class of graphs satisfies property (i'), it can not used to build efficient  ranking algorithms, since 	it has $O(N^2)$ edges. }
   Because of symmetry, we can easily see that, after
  a proper permutation of the nodes,
  $(\stackrel{\rightarrow}{\Gc}_N)_i =
  (\stackrel{\rightarrow}{\Gc}_N)_1$ for every $i = 1,\dots, N-1$. For
  the same reason, in $(\stackrel{\rightarrow}{\Gc}_N)_1$,
  $\theta_{j_1,1} = \theta_{j_2,1}$ for $j_1, j_2 = 2,\dots,N-1$. Thus,
  the only surviving edges in $(\stackrel{\rightarrow}{\Gc}_N)_1$ are
  the edges connected either to node 1 or to the reference node $N$.
  Then, the maximum   path length from node 1 to the reference is 2. Thus this family of
  graphs meets condition i) of Prop. \ref{prop:normgraphsundir}.  In
  particular, the estimate of $q_i$ is given by \[
\widehat{q}_i = \frac{2}{N} \widehat{d}_{i,N} + \frac{1}{N} \sum_{\substack{j = 1 \\ j \neq i}}^{N-1} \left( \widehat{d}_{i,j} + \widehat{d}_{j,N} \right)
\] 
\end{example}

\begin{example} \label{ex:hubs}
  Let $N'$ and $\Delta$ be any two positive numbers. Let us build the
  family of graphs $\{\Gc_N\}_{N \geq N'}$ as follows. Nodes
  $N-N'+1,\dots,N$ (a set that includes the reference) are ``hubs''
  with potentially unbounded degree. The subgraph induced by the hub
  nodes is a connected arbitrary graph. The remaining nodes are
  divided into $N'$ subsets $\Sc_1, \dots, \Sc_{N'}$. Subset $\Sc_j$,
  $j = 1,\dots,N'$, is composed of nodes with maximum degree $\Delta$,
  which are neighbors of hub node $N-j+1$ and whose other neighbors
  all belong to $\Sc_j$. It is easy to see that, for this family of
  graphs, the diameter is bounded by $N' + 1$.

  Consider a node $i \in \Sc_j$. Since all paths that reach the
  reference must pass through the hub nodes, it is easy to see that,
  in $(\stackrel{\rightarrow}{\Gc}_N)_i$, node $i$ is connected only
  to nodes belonging to $\Sc_j \cup \{N-N'+1,\dots,N\}$. Whenever the
  biased random walk on $(\stackrel{\rightarrow}{\Gc}_N)_i$ leaves
  $\Sc_j$ (by reaching hub node $N-j+1$) does not enter it any
  more. Thus, we can divide into two parts the biased random walk: the
  first on the subgraph of $(\stackrel{\rightarrow}{\Gc}_N)_i$ induced
  by $\Sc_j$, where hub node $N-j+1$ serves as reference, and the
  second on the hub nodes. Then, we can deduce the following facts.

\begin{itemize}
\item In the first part of the random walk, since hub node $N-j+1$ is
  the reference, the probability of reaching it in one step from any
  node in $\Sc_j$ is larger than $1/\Delta$. Thus, the probability of
  reaching it within $D'$ steps is upper-bounded by $\tau = 1-\left( 1 - \frac1{\Delta}\right)^{D'}$.
\item  The second part of the random walk lasts for at most $N'-1$ steps.
\end{itemize}
Thus, every node is proximal with parameters $(\tau, D'+N'-1)$, and condition ii) of Prop. \ref{prop:normgraphsundir} is satisfied. 
\end{example}

 \begin{remark}
Although 	our unweighted LS estimator is akin to the one in \cite{graphresistance}, our analysis of its performance differs substantially, also because we consider the PAC approach. Consequently, our characterization of ``good'' graphs does not coincide with that of  \cite{graphresistance}. For instance, a particular case of Example \ref{ex:hubs} is the wheel graph, which corresponds to choosing $N' = 1$ (the reference node as the only hub) and $\Delta = 3$. From \cite[Theorem 1]{graphresistance}, the wheel graph would require $W = O(N)$ comparisons per edge in order for the upper bound on the estimation error to hold, since every edge belongs to a simple path from any node to the reference. Instead, Prop. \ref{prop:normgraphsundir} allows to conclude that $W > \beta( \epsilon, \delta) \log N$ is enough to achieve the $(\epsilon, \delta)$-PAC. 
\end{remark}

\begin{example}
  Star graphs  represent a particular sequence $\{     \Gc_N\}_N$ of   acyclic
  graphs with bounded-length paths. Therefore, they satisfy
  condition i) of Prop.~\ref{prop:normgraphsundir}. In such a case, an object
  (let us say object 1) is taken as pivot  (i.e. center of the star) and qualities of all the
  other objects are estimated only through direct comparisons with the pivot.
  Observe, that, in such particular case, ranking among
  objects can be directly inferred from $\widehat{p}_{i,1}$ without
  the necessity of inverting function $F(\cdot)$.  In practice it is
  enough to rank objects according to the following rules:
  $ \widehat{r}(i) \prec \widehat{r}(j) $ iff
  $ \widehat{p}_{i,1} > \widehat{p}_{j,1}$ and
  $ \widehat{r}(i) \prec \widehat{r}(1) $ iff
  $\widehat{p}_{i,1} > 1/2$. Therefore, star graphs are appealing when
  function $F(\cdot)$ (i.e. the precise worker model) is not known.
\end{example}

\section{Results with synthetic datasets\label{sec:res_synthetic}}
\begin{figure}[t]
\centerline{\includegraphics[width=0.95\columnwidth]{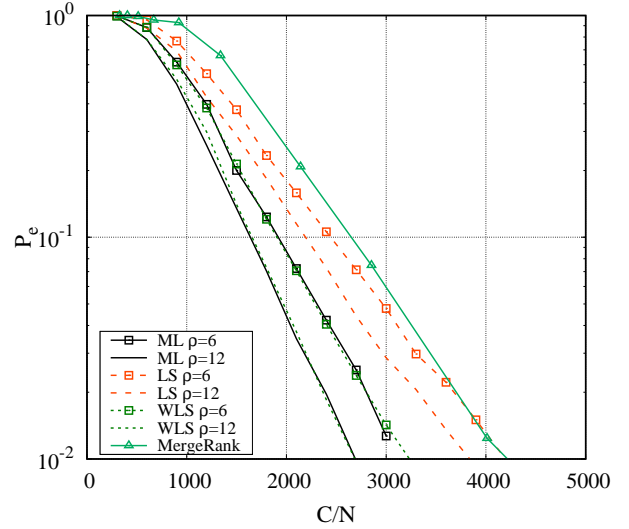}}
\caption{Error probability achieved by several ranking algorithms
  plotted versus the complexity per object $C/N$, for $N=50$.
  Object qualities are equally spaced in $[0,1)$ and the workers behave according to the Thurstone
  model.}\label{fig:figure1}
\end{figure}

\begin{figure}[t]
\centerline{\includegraphics[width=0.95\columnwidth]{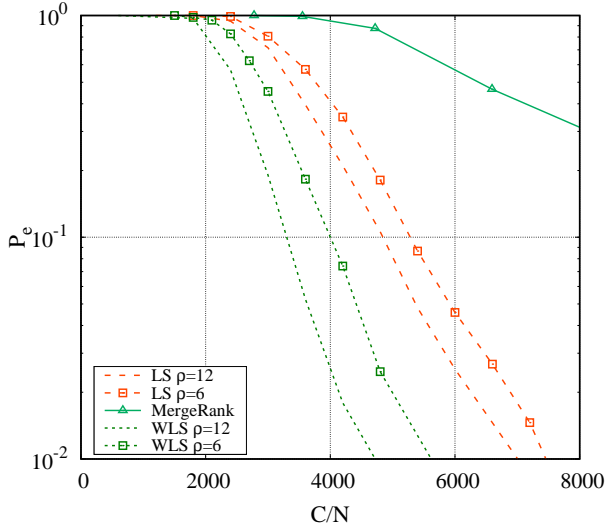}}
\caption{Error probability achieved by several ranking algorithms
  plotted versus the complexity per object $C/N$, for $N=500$.
  Object qualities are equally spaced in $[0,1)$ and the workers behave according to the Thurstone
  model.}\label{fig:figure2}
\end{figure}
We present numerical results showing the performance of our proposed
algorithms for moderate values of $N$. In Figures~\ref{fig:figure1}
and~\ref{fig:figure2} we compare the error probability achieved by
several ranking algorithms versus the complexity per object
$C/N$. Objects qualities are equally spaced in the range $[0,1)$,
i.e., object $i$ has quality $i/N$ where $N=50$
(Fig.~\ref{fig:figure1}) and $N=500$ (Fig.~\ref{fig:figure2}).
Workers' behavior is described by the Thurstone model detailed in
Section~\ref{sec:system_model} where $p_{i,j}=F(q_i-q_j)$ and
$F(\cdot)$ is the cdf of a Gaussian random variable with zero mean and
standard deviation $\sigma=0.4$.  On the $y$-axis we display the
empirical probability of generating an output which is not an
$\epsilon$-quality ranking, for $\epsilon=0.04$.  Note that an error
is counted whenever at least two objects, whose quality difference
exceeds $\epsilon$, appear swapped in the estimated ranking.  The
curve labeled ``MergeRank'' refers to the {\em Merge-Rank} algorithm
proposed in~\cite{falahatgar2017fewassumptions}, which we consider as
a performance reference.  The LS, WLS~\footnote{Reported WLS results have been obtained by setting $\chi=10^{-4}$}, and ML algorithms have been applied to
randomly generated regular graphs~\cite{bollobas2001random} whose
nodes have degree $\rho=6$ (lines with square marker) and $\rho=12$
(lines without markers).  The figure shows the superior performance of
our ranking algorithm. It is interesting to observe that the WLS
algorithm provides significant enhancements with respect to the LS
algorithm and almost perfectly matches the performance of the more
(computationally) complex ML approach. As the number of nodes
increases our proposed solutions substantially outperform the
``MergeRank'' algorithm.
\begin{figure}[t]
\centerline{\includegraphics[width=0.95\columnwidth]{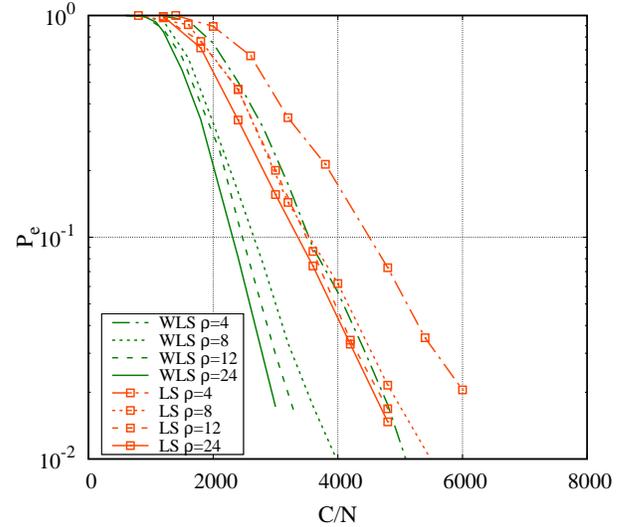}  }
\caption{Performance of the LS and WLS ranking algorithms plotted
  against the complexity per object $C/N$, for $N=200$ objects. Object qualities
  are drawn from a uniform distribution in $[0,1)$ and the workers
  behave according to the Thurstone model.}\label{fig:figure3}
\end{figure}

Figure~\ref{fig:figure3} compares the performance of the LS and of the
WLS algorithms for $N=200$ objects. Object qualities are randomly
generated according to a uniform distribution in $[0,1)$. Other system
parameters are set as in Figure~\ref{fig:figure2}. The figure
reports the empirical error probability plotted versus the number of
tests per node, $C/N$, for different values of the degree of the nodes
in the graph.  We first observe that, given $C/N$, the number of tests
per edge of the graph decreases as the degree, $\rho$,
increases. Hence, as $\rho$ increases, distances between pairs of
nodes (corresponding to edges of the graph) are estimated with a
decreasing accuracy. In spite of that, a larger number of neighbors
for each node (i.e., a larger $\rho$) leads a more reliable estimation
of object qualities. This effect is more evident when the WLS
algorithm is employed. Indeed, because of the weights $\omega_{i,j}$,
as $\rho$ increases, WLS is able to well exploit the increasing number
of highly-reliable edges in the graph connecting objects with similar
qualities; at the same time WLS is able to limit the impact of the
greater number of scarcely-reliable edges that connect objects with
largely different qualities.

\subsection{Adaptive multistage approach}
The performance of the proposed ranking algorithms can be improved by
adopting a multistage approach where, at each stage, new edges are
added to the graph, depending on the quality estimates obtained at
previous stage. The rationale of this approach stems from the fact
that such algorithms provide approximate rankings,
in which  the probability  of swapping the order of two  objects increases as
their distance (in terms of their qualities) decreases.  Therefore, in
order to mitigate  this phenomenon and, thus, improve the reliability of
the estimate, it is convenient to (i) add to the graph extra edges 
connecting neighboring objects (in terms of their estimated
qualities); (ii) assign additional workers to the already existing
edges connecting the aforementioned neighboring objects.  This
procedure can be iterated until a desired performance level is
achieved.

\begin{figure}[t]
\centerline{\includegraphics[width=0.95\columnwidth]{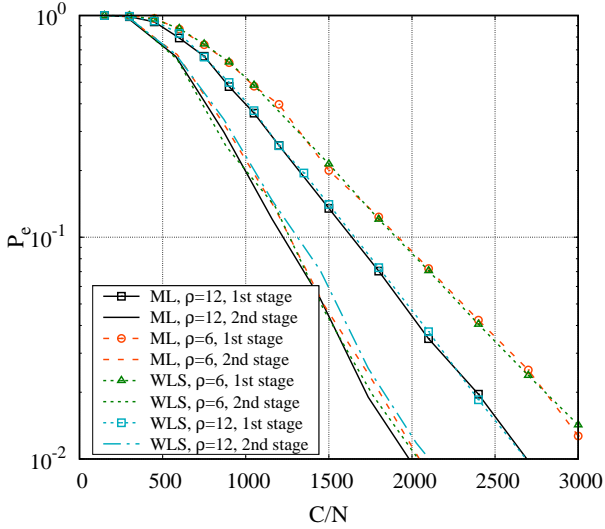}  }
\caption{Error probability provided by ML and WLS algorithms when a 2-stage adaptive approach is employed, for
$\rho = \rho^{(1)}=\rho^{(2)}=6,12$, and $N=50$.}\label{fig:figure4}
\end{figure}

In our simulation setup, we have considered a 2-stage approach where
we first apply the estimation algorithm to a random regular graph,
$\Gc^{(1)}(\Vc,\Ec^{(1)})$, of degree $\rho^{(1)}$, obtaining the
vector of estimates $\widehat{\qv}^{(1)}$.  In the second stage, we
create a new regular graph, $\Gc^{(2)}(\Vc,\Ec^{(2)})$ of degree
$\rho^{(2)}$, where each node is connected to its 
$\rho^{(2)}$ closest neighbors, according to the estimates $\widehat{\qv}^{(1)}$. Finally, the estimation algorithm is applied
to the graph $\Gc^{(1)}\cup \Gc^{(2)}$ obtaining the output
$\widehat{\qv}^{(2)}$ which is used to infer the ranking.  In
Figure~\ref{fig:figure4} we show the performance of the ML and WLS
algorithms when the proposed multistage approach is employed. For both
algorithms we show the error probability versus the number of tests per
object, $C/N$, for $\rho = \rho^{(1)}=\rho^{(2)}=6,12$, and $N=50$.
We observe that the second stage allows for a significant improvement
of the performance and a reduction of about 60\% of the required tests
per object for $\rho=6$ and of about 30\% for $\rho=12$.  In both
cases the performance of the WLS algorithm is very close to that
provided by the ML algorithm.

\section{Results with real-world datasets\label{sec:results_real_world}}

In this section, we show that our algorithm works well even when
considering a real scenario, where the ``evaluations'' are the outcome
of experiments, and not synthetically generated by simulations. In
particular, we consider five recent seasons of the English Premier
League and build up a $N=20$ complete graph, where nodes are the
football teams and edges are the matches between each pair of
them. The match between team $i$ and team $j$ is considered as lasting
for 180 minutes, since it includes both the round when $i$ is at home
and the round where $i$ is away. If team $i$ has scored $x_{ij}$ goals
in the match against team $j$, we count
$K_{ij} = \alpha x_{ij} + \beta$ evaluations in favor of $i$ when
compared to $j$, where $\alpha > 0$ and $\beta \geq 0$ are
constant. The total number of comparisons between $i$ and $j$ is then
simply $W_{ij} = K_{ij} + K_{ji}$\footnote{With this definition, the
  edge between $i$ and $j$ may be actually missing if
  $x_{ij} = x_{ji} = 0$ and $\beta = 0$.}.

The WLS algorithm has been run with $\chi=10^{-4}$ and both the Thurstone 
and BTL models, to see the influence of the underlying
worker model. The true ranking is assumed to be the final season
ranking. The results have been plotted in terms of the Kendall tau
distance, which counts the number of inversions in the estimated
ranking with respect to the true ranking, i.e. the number of pairs
$(i,j)$ for which $i$ is ranked better than $j$ in the true ranking
and worse than $j$ in the estimated one.

\begin{figure}[t]
\centerline{\includegraphics[width=0.95\columnwidth]{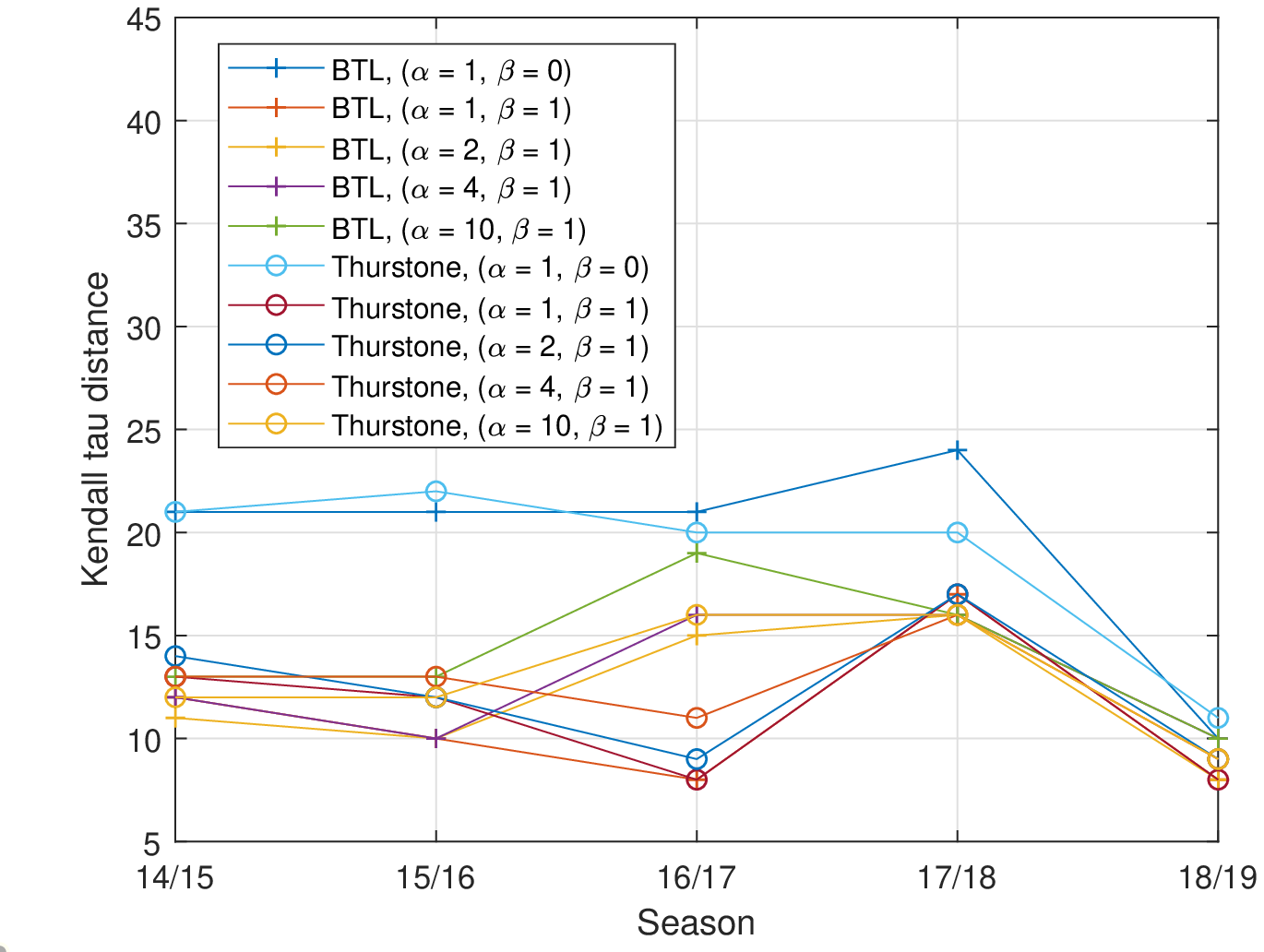}  }
\caption{Distance between true and estimated ranking for Premier League scores. The $x$-axis is the season. The $y$-axis is the Kendall tau distance (number of inversions) between the final season ranking and the output of the WLS algorithm, for different choices of the model and of the parameters.}\label{fig:figureFB}
\end{figure}

Results are shown in Fig. \ref{fig:figureFB}. First, we can observe
that the performance is better with $\beta > 0$ than with $\beta = 0$,
since in the latter case there might be some edges for which the
estimated preference probability is very close to either 0 or 1. Such edges are
automatically dropped by the WLS algorithm, while in the former case
each object in each comparison receives at least $\beta$ preferences,
so that all edges are used for ranking computation. Second, the
Thurstone model seems to be slightly better suited than the
BTL model. Third, while in most cases, the influence of
parameters is limited, there are cases (like season 16/17) that are
more sensible to the choice of $\alpha$ and $\beta$. It is worth mentioning that, in \cite{DBLP:journals/corr/Cucuringu15}, the Sync-Rank algorithm is applied to older Seasons of the Premier League. Comparatively, for $\beta >0$, Kendall tau distance for our algorithm never goes
beyond 20, giving rise to a Kendall correlation larger than 0.90, which is a
better result than those shown in
\cite{DBLP:journals/corr/Cucuringu15}.


\section{Conclusions\label{sec:conclusions}}
In this paper, we focused on the problem of ranking $N$ objects
starting from a set of noisy pairwise comparisons. Objects are assumed
to be endowed with intrinsic qualities. The probability $p_{i,j}$ that
object $i$ is preferred to $j$ is given by an arbitrary smooth
monotonic function of the difference between qualities of the two
competitors.  For such a scenario we developed a class of
order-optimal ranking algorithms, i.e. algorithms, which are provably
$(\epsilon, \delta)$-PAC when
$O(\frac{N}{\epsilon^2}\log( \frac{N}{\delta}))$ comparisons are
blindly allocated in a single round.  Our ranking procedure is based
on the reconstruction of object qualities, from pairwise quality
differences, by adopting a simple LS approach. The analysis
establishes a parallelism between the quality estimations process and
the cumulative reward accumulated by random walks on graphs. Finally,
by simulation, we show that the performance of our algorithms, and
further variants, is extremely good also in non-asymptotic scenarios
and approaches that obtained by the ML algorithm.  Our results
complement and extend previous recent studies
\cite{szorenyi2015online,falahatgar2017maximum,falahatgar2018limits}
on the minimal complexity of ranking algorithms under different
non-parametric preference models.

\bibliographystyle{IEEEtran}
\bibliography{refs}

\newpage
\setcounter{page}{1}
\onecolumn
\begin{center}
  {\Huge Ranking a set of objects: a graph based \\ least-square approach} \\
  \medskip
   {\Large Supplemental material}
\end{center}

\appendices

\section{Proof of Proposition \ref{second-mom}
	\label{app:second-mom}}
Given that:
\begin{equation} \label{eq:final1}
\widehat{\qv}  = \Mm^{-1} (\Hm \odot \widehat{\Dm}) \onev = \qv +  \Mm^{-1} (\Hm \odot \Zm) \onev 
\end{equation} 	
 we can write the MSE, on the estimate of $\qv$ as
\begin{eqnarray} \label{eq:MSE_bound}
  {\rm MSE}
  &=&\EE[(\widehat{\qv}-\qv)\Tran (\widehat{\qv}-\qv) ] \non
  &=& \EE[\onev\Tran (\Hm\odot \Zm)\Tran \Cm (\Hm\odot \Zm)\onev]\non
  &\le &  \lambda_\Cm^{\rm max}\EE[\onev\Tran (\Hm\odot \Zm)\Tran(\Hm\odot\Zm)\onev]
\end{eqnarray}
where $\Cm = (\Mm^{-1})\Tran \Mm^{-1}$ is a deterministic matrix and $\lambda_\Cm^{\rm max}$ is the largest eigenvalue of $\Cm$. By using
the definition of the matrix $\Hm$ the term $\EE[\onev\Tran (\Hm\odot \Zm)\Tran (\Hm\odot \Zm)\onev]$ in~\eqref{eq:MSE_bound} can be expanded as
\begin{eqnarray}\label{eq:secmom-z}
  &&\hspace{-4ex}\EE[\onev\Tran (\Hm\odot \Zm)\Tran (\Hm\odot \Zm)\onev] \non
  &\mathord{=}&\hspace{-2ex}\sum_{i\in \Nc_k} \sum_{j\in \Nc_k} \sum_k \frac{1}{\rho_k^2}\EE\left[ z_{k,i}z_{k,j} \right]\non
  &\mathord{=}&\hspace{-2ex}\sum_k\frac{1}{\rho_k^2}\sum_{i\in \Nc_k} \EE\left[z^2_{k,i}\right]\mathord{+} \sum_k \frac{1}{\rho_k^2}\sum_{i\in \Nc_k}\sum_{j\in \Nc_k, j\neq i}\hspace{-2ex} \EE\left[ z_{k,i}\right] \EE\left[z_{k,j}\right]\,.\non    
\end{eqnarray}  
In order to evaluate the averages in~\eqref{eq:secmom-z} we consider
the following linear approximation of $z_{i,j}$
\[ z_{i,j} = \gamma_{i,j}y_{i,j} +O(y_{i,j}^2) \]
where $\gamma_{i,j}=\frac{\dd F^{-1}(p)}{\dd p}|_{p=p_{i,j}}$ are positive constants. By applying this result
to~\eqref{eq:secmom-z} we obtain
\begin{equation}
  \EE[z_{i,j}]= \gamma_{i,j}\EE[y_{i,j}] +O(E[y_{i,j}^2]) = O\left(\frac{ p_{i,j}(1-p_{i,j})}{W}\right)
\end{equation}
thanks to the fact that $\EE[y_{i,j}]=0$ and that
$\EE\left[ y^2_{i,j}\right] = \frac{p_{i,j}(1-p_{i,j})}{W}$. As for
the term $\EE\left[z^2_{k,i}\right]$ we can write
\begin{eqnarray}
  z_{i,j}^2
  &=& |z_{i,j}|^2 \non
  &=& \left|\gamma_{i,j}y_{i,j}+O\left(y^2_{i,j}\right)\right|^2 \non
  &\le& \gamma^2_{i,j}y_{i,j}^2 +O\left(y^4_{i,j}\right)+2\gamma_{i,j}O\left(|y^3_{i,j}|\right)
\end{eqnarray}
Since, $y_{i,j}$ is defined as a difference of two probabilities we have $|y_{i,j}|\le 1$. Then $y^4_{i,j}\le |y^3_{i,j}|\le  y^2_{i,j}$.
This result allows to upper-bound the term $\EE[z_{i,j}^2]$ as
\begin{equation}
  \EE[z_{i,j}^2] \le \EE\left[\gamma^2_{i,j}y_{i,j}^2 +O\left(y^2_{i,j}\right)\right] \le c_1\EE[y_{i,j}^2]
\end{equation}
for some constant $c_1>0$. In conclusion the MSE can be bounded as
\begin{eqnarray}
  &&\hspace{-4ex}{\rm MSE}\non
  &\le & \lambda_\Cm^{\rm max}\EE[\onev\Tran (\Hm\odot \Zm)\Tran(\Hm\odot \Zm)\onev] \non
  &=& \lambda_\Cm^{\rm max}\left[\sum_k\sum_{i\in \Nc_k} \frac{\EE\left[z^2_{k,i}\right]}{\rho_k^2}+ \sum_k \sum_{i\in \Nc_k}\sum_{\substack{j\in \Nc_k\\j\neq i}}\frac{\EE\left[ z_{k,i}\right] \EE\left[z_{k,j}\right]}{\rho_k^2} \right]\non
  &\le &  \lambda_\Cm^{\rm max}\left[c \sum_k\sum_{i\in \Nc_k}\frac{O\left(\frac{ p_{k,i}(1-p_{k,i})}{W}\right)}{\rho_k^2}+ \sum_k \sum_{i\in \Nc_k}\sum_{\substack{j\in \Nc_k\\j\neq i}} \frac{O\left(\frac{1}{W^2}\right)}{\rho_k^2}\right]\non
  &\le&  \frac{c_1}{4}\lambda_\Cm^{\rm max}\sum_k\frac{1}{\rho_k^2}\sum_{i\in \Nc_k} O\left(\frac{1}{W}\right)\non
  &\le&  c \frac{N\lambda_\Cm^{\rm max}}{W\rho_{\inf}}
\end{eqnarray}
where, $c$ is a constant, we assumed that $\rho_i$ is uniformly upper-bounded for any $i$, we defined $\rho_{\inf}\triangleq \inf_i\rho_i$,
and used the bound $p_{i,j}(1-p_{i,j})\le \frac{1}{4}$.
\section{Proof of Proposition  \ref{EPAC}
	\label{app:EPAC}}
From~\eqref{eq:final} we can write the error on the quality estimates as $\widehat{\qv} - \qv = \Mm^{-1}(\Hm\odot \Zm)\onev$.
We can then upper-bound the term $\sup_i|\widehat{q}_i-q_i|$ by using the infinity norm as follows
\begin{eqnarray}
  \sup_i|\widehat{q}_i-q_i|
  &=& \|\widehat{\qv}-\qv\|_\infty\non
  &=& \|\Mm^{-1}(\Hm\odot \Zm)\onev\|_\infty\non
  &\le& \|\Mm^{-1}\|_\infty\|(\Hm\odot \Zm)\onev\|_\infty\non
  &\le& \|\Mm^{-1}\|_\infty\sup_i \left|\sum_{j\in\Nc_i}\frac{z_{i,j}}{\rho_i}\right|\non
    &\le& \|\Mm^{-1}\|_\infty\sup_i \sup_{j\in\Nc_i}|z_{i,j}|\non
    &=& \|\Mm^{-1}\|_\infty\sup_{(i,j)\in\Ec}|z_{i,j}|
\end{eqnarray}
thanks to the fact that $\Hm$ is substochastic. Let $\limsup_{N\to \infty}\|\Mm^{-1}\|_\infty=K<\infty$. Then we can write
\begin{eqnarray}
  \PP\left(\sup_i|\widehat{q}_i-q_i|>\epsilon\right)
  &\le&  \PP\left(\|\Mm^{-1}\|_\infty\sup_{(i,j)\in\Ec}|z_{i,j}|>\epsilon\right)\non
  &\le&  \PP\left(\sup_{(i,j)\in\Ec}|z_{i,j}|>\frac{\epsilon}{K}\right)     
\end{eqnarray}
which converges to 0 as $N\to \infty$ under the conditions ii) and iii), as stated in
Appendix~\ref{app:additional}, Proposition~\ref{prop-YZ}.

The last statement of the proposition can be proved by considering a sequence of operators
$\Ac\{\cdot\}:\RR^{N\times N} \to \RR^{N}$ mapping the sequence of
matrices $\Am=\{a_{i,j}\}$ with bounded norm
$\|\Am\|_{\infty}=\sup_{i,j}|a_{i,j}|\le 1$ into a set of vectors
$\Mm^{-1}(\Hm\odot\Am)\onev$ with uniformly bounded $\infty$-norm. Let  $K\ge \|\Mm^{-1}(\Hm\odot \Am)\onev\|_\infty$. 
We know that, as $N\to \infty$, $\sup_{(i,j)\in\Ec}|z_{i,j}|<\frac{\epsilon}{K}$  with a probability larger than $1-\delta$.
Then, we can write $\Zm=\kappa \Am$ where $\kappa=\|\Zm\|_\infty$ and $\Am=\frac{\Zm}{\|\Zm\|_\infty}$
is such that $\|\Am\|_{\infty}=1$.
It follows that $\|\Mm^{-1}(\Hm\odot \Zm)\onev\|_\infty=\kappa\|\Mm^{-1}(\Hm\odot \Am)\onev\|_\infty<\epsilon$
 with a probability larger than $1-\delta$. 

\section{Proof of Proposition \ref{prop:normgraphsdir}
	\label{app:normgraphsdir}}
Let us consider the directed graph $\widehat{\Gc}$ on $N$ nodes.  The
proof of the proposition descends from the fact that~\eqref{eq:syst_eq_dir} 
provides an explicit expression for operator
$\bold b={\bold M}^{-1}(\bold H \odot \bold{A} ) \bold{1} $, i.e.,
\[
b_i= \frac{1}{\rho^-_i}  \sum_{j_1\in\Nc^-_i} \left(  a_{j_1i}+ \frac{1}{\rho^-_{j_1}}\sum_{j_2\in\Nc^-_{j_1}}\tilde{a}_{j_2} \right)
\]
where
\[ \tilde{a}_{j_2} =  a_{j_2j_1}+ \ldots \frac{1}{\rho^-_{j_{k_{i}-1}}}\sum_{j_{k_i}\in\Nc^-_{j_{k_{i}-1}}} a_{j_{k_{i}-1} j_{l_{i}}} \]
and  $l_i$ is the length of the longest (simple) path from $i$ to the
reference node $n$.  Now, we denote with $\Pc_i$ the set of
paths from $i$ to $n$, and for any path $p\in \Pc_i $ we
denote with $L(p)$ and $p(h)$ respectively the graph-theoretical
length (expressed in number of hops) of $p$ and $h$-th node along $p$.
By inspection, it can be easily seen that the previous expression can
be rewritten as follows: \beq \label{eq:explicit_unbiased} b_i =
\sum_{p\in\mathcal{P}_i} \prod_{h=1}^{L(p)} \frac{1}{\rho_{p(h)}^-}
\sum_{h\in 1}^{L(p)} a_{p(h) p(h+1)} \eeq From the above expression,
it is rather immediate to check that
$ \|b_i\|_{\infty}< \overline{D}_{\widehat \Gc} \sup_{i,j}| a_{i,j}|$
where $\overline D_{\widehat \Gc}$ is the length of the longest path
to node $n$ on $\widehat \Gc$. Thus, if $\overline D_{\widehat \Gc}$
is finite and $\sup_{i,j}|a_{i,j}|= 1$, also $\|b_i\|_{\infty}$ is
finite, and condition i) of Proposition \ref{prop:normgraphsdir}
 is demonstrated.

Observe that, if we interpret $a_{p(h) p(h+1)}$ as the elementary
reward associated to edge $(p(h), p(h+1))$,
$\sum_{h\in 1}^{L(p)} a_{p(h) p(h+1)} $ can be regarded as the total
reward associated to path $p$.  Furthermore,
$\prod_{h=1}^{L(p)}\frac{1}{\rho_{p(h)}^-}$ is equal to the
probability, for a Random Walker (henceforth, RWer) on
${\widehat \Gc}$ that starts in $i$ and ends in $n$, to take path $p$.
As a result, $b_i$ can be interpreted as the expected total reward
accumulated by the RWer starting in $i$ and stopping as soon as it
reaches node $n$.

Now, assuming $\sup_{i,j}|a_{i,j}|= 1$, we have that, for $i = 1,\dots,n$,
\begin{eqnarray}
  |b_i|
  &\le& \sum_{p\in\Pc_i} L(p)\prod_{h=1}^{L(p)}\frac{1}{\rho_{p(h)}^-} \non
  &=& \sum_{t=1}^{\infty} \PP \{\text{RWer is still active after $t$ hops} \} \non
  &=& \EE[T_i]
\end{eqnarray}
where $\EE[T_i]$ represents the average stopping (i.e., absorbing) time for the RWer.
Therefore, from the above considerations we can write 
$\limsup_{N\to \infty} \sup_i |b_i|<\infty$ if
$\limsup_{N\to \infty} \sup_{i} \EE[T_i]<\infty $.  Now, we
show that under assumptions ii)
$\limsup_{N\to \infty} \sup_{i} \EE[T_i]<\infty$.

Let $D$ denote the diameter of $\widehat \Gc$.  By construction, from
any node $i < n$, there exists one path of length at most $D$ that
leads to node $n$.  We call the shortest path from $i$ to $n$ as
\emph{critical}. Define the following events:
\begin{eqnarray}
\Ac[k] &=& \{ \text{RWer is still active after $t = kD$ hops}\} \non
    \Wc_i[k] &=& \{ \text{ RWer in $i$ (active) at time $t = kD$ }\} \non
 \Kc_i[k] &= & \{ \text{RWer takes the critical  path  from $i$}\non  
 && 	\text{at $t = kD$ and follows it up to  its end}\}                       
\end{eqnarray}
We have
\begin{eqnarray}
  &&\hspace{-6ex}\PP \left\{\Ac[k] \mid \Ac[k-1] \right\}\non
  &\le& 
 \sum_{i=1}^{n-1} \left( 1- \mathbb{P} \left\{\Kc_i[k-1]  \right\} \right)  \mathbb{P} \left\{\Wc_i[k-1]  \mid \Ac[k-1] \right\}
\end{eqnarray}
Now, uniformly  over $i$,
\[
 \mathbb{P}\left\{\Kc_i[k-1]\right\} \ge \left(\frac{1}{\sup_i \rho_i^-}\right)^D 
\]
 Therefore, since by construction $ \sum_{i=1}^{n-1} \mathbb{P} \left\{\Wc_i[k-1]  \mid \Ac[k-1] \right\} =1 $
 we have that:
 \begin{align*}
 & \mathbb{P} \left\{\Ac[k] \mid \Ac[k-1] \right\} \le 1 - \left(\frac{1}{\sup_i \rho_i^-}\right)^D 
 \end{align*}
from which we get:
\[
 \mathbb{P} \left\{\Ac[k] \right\} = \prod_{k'=1}^k \mathbb{P} \left\{\Ac[k'] \mid \Ac[k'-1] \right\} \le  \left[1-\left(\frac{1}{\sup_i \rho_i^-}\right)^D\right]^{k}. 
\]
The assertion concerning condition ii) follows immediately,

Now, the previous argument can be easily extended under iii).  Let $D$
be now any finite integer sufficiently large and keep the same
definitions of the events $\Ac[k]$, $\Wc_i[k]$ and $\Kc_i[k]$. We
denote with $\Bc$ the set of proximal nodes, i.e., of the nodes
satisfying the $(\tau, D)$ property. Let also
$\overline{\Bc} = \Vc \backslash \{\Bc \cup \{N\}\}$ be the set of
non-proximal nodes.  We qualify as critical any path that, from a
given node, reaches the reference node within $D'$ hops.  In
particular let $\alpha>0$ be a uniform lower bound to the fraction of
in-edges connecting an arbitrary node $v$ to $\Bc$.

Note that, by construction, at every instant  the RWer, if active,  is visiting a node in $\Bc$ with a probability at least $\alpha$.
Therefore:
\begin{eqnarray}
  &&\hspace{-4ex}\PP \left\{\Ac[k] \mid \Ac[k-1] \right\}\non
  &\le & \sum_{i=1}^{N-1} \left( 1- \PP\left\{\Kc_i[k-1]  \right\} \right)  \PP\left\{\Wc_i[k-1]  \mid \Ac[k-1] \right\}  \\ 
  & = & (1- \tau)\sum_{i\in \Bc}\PP \left\{\Wc_i[k-1]  \mid \Ac[k-1] \right\}\non
        &&  \qquad +  \sum_{i\in \overline{\Bc}} \PP\left\{\Wc_i[k-1]  \mid \Ac[k-1] \right\} \\
& \le &  (1-\tau) \alpha+ (1- \alpha) 
\end{eqnarray}
Then, proceeding as before we get the result.

\section{Proof of Proposition \ref{prop:equivdirundir} \label{app-prop:equivdirundir}
}

The proof descends from the analysis of \eqref{eq:final}. Let us define
\beq
\Hm = \left[
\begin{array}{c|c}
\Tm & \vvv \\
\hline
0 \dots 0 & 0 
\end{array}
\right]
\eeq
so that 
\beq
\Mm^{-1} = (\Id_{n} - \Hm)^{-1} = \left[
\begin{array}{c|c}
(\Id_{n-1}-\Tm)^{-1} & (\Id_{n-1}-\Tm)^{-1}\vvv \\
\hline
0 \dots 0 & 1 
\end{array}
\right]
\eeq
Thus, from \eqref{eq:final}, we get
\begin{eqnarray} \label{eq:rwthroughundir}
\widehat{q}_i &=& \sum_{j=1}^{n-1} \sum_{\ell=1}^{n-1} ((\Id-\Tm)^{-1})_{i\ell} (\Tm \odot \widehat{\Dm}_{1:N-1,1:N-1})_{\ell j} \non
&& + \sum_{\ell=1}^{n-1} ((\Id-\Tm)^{-1})_{i,\ell} (\vvv \odot \widehat{\Dm}_{1:N-1,N})_{\ell} \non
& = & \sum_{j=1}^{N-1} \sum_{\ell \in \Nc_j} \theta_{\ell,i} \widehat{d}_{\ell,j} + \sum_{\ell \in \Nc_n} \theta_{\ell,i} \widehat{d}_{\ell, n} \non
& = & \sum_{\substack{j, \ell :  (j, \ell) \in \Ec, \\ \theta_{\ell, i} > \theta_{j,i}}} (\theta_{\ell,i}-\theta_{j,i}) \widehat{d}_{\ell,j} 
\end{eqnarray}
where $\theta_{j,i}$ is defined in~\eqref{eq:theta_def} 
for $j < n$, $\theta_{N,i} = 0$,  and we have used the fact that $\widehat{d}_{\ell,j} = -\widehat{d}_{j,\ell }$. 

Now consider the directed graph
$\stackrel{\rightarrow}{\Gc}_i$. Without loss of generality, let us
consider the case $i=1$ and, whenever possible, let us drop the
subscript and write simply $\stackrel{\rightarrow}{\Gc}$, $\theta_j$,
etc. Let $\Nc_j^-$ and $\Nc_j^+$ be the out-neighbors and in-neighbors
of node $j$, respectively. We first prove the following lemma.

\begin{lemma} \label{lemma:inandout}
With the previous definitions,
\beq  \label{eq:inandout1}
\sum_{\ell \in \Nc_j^-} (\theta_{j }-\theta_{\ell }) = \left\{ 
\begin{array}{cc}
1, & j = 1, \\
\sum_{\ell \in \Nc_j^+} (\theta_{\ell }-\theta_{j }), & 1<j<N.
\end{array} \right.
\eeq
\end{lemma}
\pf
For $1<j<N$, we have on the indirect graph $\Gc$
\begin{eqnarray} \label{eq:inandout2}
\sum_{\ell \in \Nc_j} \theta_{j} & = &
((\Id-\Tm)^{-1})_{1 j} \non
& = & ((\Id-\Tm)^{-1}- \Id)_{1 j} \non
& =  & (\Tm + \Tm^2 + \Tm^3 + \dots)_{1 j} \non
& =  & ((\Id + \Tm + \Tm^2 + \dots) \Tm)_{1 j} \non
& =  & \sum_{\ell = 1}^{N-1} ((\Id-\Tm)^{-1})_{1 \ell} T_{\ell j} \non
 & = & \sum_{\ell \in \Nc_j} \theta_{\ell} 
\end{eqnarray}

Now, from the definition of $\stackrel{\rightarrow}{\Ec}$, it turns
out that $\Nc_j = \Nc_j^- \cup \Nc_j^+ \cup \Nc_j^0$, where $\Nc_j^0$
is the subset of neighbors $\ell$ of $j$ such that
$\theta_j =\theta_{\ell}$. Reordering the terms in
\eqref{eq:inandout2}, we obtain \eqref{eq:inandout1}. For $j= 1$, we
can proceed in the same way, with the additional remark that
$\Nc_1^{+}$ is empty.  \qedsymb

Consider a perturbed Random Walker (RWer), which starts from node 1 and, given that it has reached node $j$, proceeds through edge $(j,\ell) \in \stackrel{\rightarrow}{\Ec}$ with probability $\eta_{j \rightarrow \ell}$ defined in~\eqref{eq:bias_prob}. 
The proof of Proposition~\ref{prop:equivdirundir} 
consists in showing that the probability for the RWer to pass through edge $(j,\ell) \in \stackrel{\rightarrow}{\Ec}$ is equal to $\theta_{j }-\theta_{\ell }$. Let us define, for $j < N$
\beq
\xi_j = \Pr\{\mbox{RWer passes through node $j$}\}
\eeq
We know by definition that $\xi_1 = 1$. Moreover, for $\ell > 1$, $\xi_{\ell}$ satisfies the linear equation
\beq \label{eq:node_prob}
\xi_{\ell} = \sum_{(j,\ell) \in \stackrel{\rightarrow}{\Ec}} \xi_j \eta_{j \rightarrow \ell} 
\eeq
Let $\stackrel{\rightarrow}{\Hm}$ be the random-walk Laplacian matrix for the perturbed random walk defined above,
i.e., $(\stackrel{\rightarrow}{\Hm})_{j\ell} = p_{j \rightarrow \ell}$. Let also $\stackrel{\rightarrow}{\Tm} = \stackrel{\rightarrow}{\Hm}_{1:N-1,1:N-1}$. Finally, let $\xiv = (\xi_2,\dots,\xi_{N-1})$ be the row vector of node probabilities. Now, \eqref{eq:node_prob} can be written in matrix form as
\beq
\xiv = \xiv \stackrel{\rightarrow}{\Tm}_{2:N-1,2:N-1} + \stackrel{\rightarrow}{\Tm}_{1,2:N-1}
\eeq  
which can be solved univocally as
\beq \label{eq:node_prob_sol}
\xiv = \stackrel{\rightarrow}{\Tm}_{1,2:N-1} \left(\Id - \stackrel{\rightarrow}{\Tm}_{2:N-1,2:N-1} \right)^{-1}
\eeq
Now notice also that $\xi_j^* = \sum_{\ell \in \Nc_j^+} (\theta_{\ell }-\theta_{j })$ satisfies \eqref{eq:node_prob}, thanks to the definition of $\eta_{j \rightarrow \ell}$ in~\eqref{eq:bias_prob} 
and Lemma \ref{lemma:inandout}. Then, since \eqref{eq:node_prob} has a unique solution given by \eqref{eq:node_prob_sol}, we conclude that $\xiv = \xiv^* = (\xi_2^*,\dots, \xi_{N-1}^*)$.

As a consequence, the probability for the RWer to pass through edge $(j,\ell) \in \stackrel{\rightarrow}{\Ec}$ will be given by
\[
\xi_j^* \eta_{j \rightarrow \ell} = \theta_{j}-\theta_{\ell }
\] 
Thus, the average total reward accumulated by the RWer before being absorbed will be given by 
\beq
\check{q}_1 = \sum_{j, \ell :  (j, \ell) \in \stackrel{\rightarrow}{\Ec}} (\theta_{\ell i}-\theta_{j i}) \widehat{d}_{\ell j}
\eeq
which coincides with \eqref{eq:rwthroughundir}. This concludes the proof.

\section{Proof of Proposition \ref{prop:normgraphsundir}\label{app-prop:normgraphsundir}
}

The proof descends from the fact that~\eqref{eq:syst_eq_dir} provides
an explicit expression for the operator
$\bold b={\bold M}^{-1}(\bold H \odot \bold{A} ) \bold{1} $:
\beq \label{eq:explicit_biased} b_i = \sum_{\pi \in \Pc_i}\prod_{h =
  1}^{L(\pi)} \eta_{\pi(h-1) \rightarrow \pi(h), i} \sum_{h =
  1}^{L(p)} a_{\pi(h-1),\pi(h)} \eeq where $\mathcal{P}_i$ is the set
of paths from $i$ to $n$, and, for any path $\pi\in \mathcal{P}_i $,
$L(\pi)$ and $\pi(h)$ are the graph-theoretical length (expressed in
number of hops) of $\pi$ and $h$-th node along $\pi$,
respectively. Notice that \eqref{eq:explicit_biased} is similar to
\eqref{eq:explicit_unbiased}, except for the expression of the
probabilities associated to the graph edges.

Notice that conditions i) and ii) of  Prop. 3.5 
coincide with conditions i) and iii) of Prop. 3.3.  
 The proof of Prop. 3.5 
 then follows immediately from the fact that the proofs of conditions i) and iii) of Prop. 3.3 
  do not explicitly depend on the expression of the edge probabilities.

\section{Additional results\label{app:additional}}
\begin{proposition}\label{prop-YZ}
For any $ \epsilon>0$ and $\delta>0$, there exists $\beta(\epsilon,\delta)$ such that, as $N\to \infty$, 		
\begin{equation}\label{eq:prop-YZ}
  \PP\left(\sup_{(i,j) \in \Ec}|y_{i,j}|>\epsilon\right) < \delta  \quad\mbox{ and }\quad \PP\left(\sup_{(i,j) \in \Ec}|z_{i,j}|>\epsilon\right) < \delta
\end{equation}
provided that $W > \beta(\epsilon, \delta) \log N $   with $ \beta(\epsilon, \delta)= O\left(\frac{1}{\epsilon^2} \frac{ \log \frac{N}{\delta} } {\log N} \right) $
 and the total number of edges of graph $\Gc$ is $|\Ec| =O(N)$.
\end{proposition}

\pf
We first use the union bound and write\\ $\PP\left(\sup_{(i,j) \in \Ec}|y_{i,j}|>\epsilon\right)\le \sum_i\PP(y_{i,j}>\epsilon)+\sum_i\PP(-y_{i,j}>\epsilon)$.
We then observe that the MGF $\phi_{y_{i,j}}(t)$ of $y_{i,j}$ is given by:
\[
\phi_{y_{i,j}}(t)= \ee^{-tp_{i,j}} \left(1+ p_{i,j}(\ee^{\frac{t}{W}}-1)\right)^W\,. 
\]
Then we bound  $\PP(y_{i,j}>\epsilon)$ by applying  the Chernoff bound:
\[
 \PP(y_{i,j}>\epsilon) \le \inf_{t>0} \frac{\phi_{v_i}(t)}{\ee^{\epsilon t}} \le  \frac{\phi_{v_i}(t)}{\ee^{\epsilon t}}
\]
 By setting $t=\zeta \log N $, and  $W=\beta \log N$, for a sufficiently large $\beta=\beta(\epsilon. \delta)$ ,
we have 
$$\PP(y_{i,j}>\epsilon) \le \exp\left(\left[-\zeta p_{i,j}  + \beta \log(  1+ p_{i,j}(\ee^{\frac{\zeta}{\beta}}-1)\right] \log N \right), $$
with $ \beta \log(  1+ p_{i,j}(\ee^{\frac{\zeta}{\beta}}-1)= \beta ( \log (1 +p_{i,j}\frac{\zeta}{\beta}+ O(\frac{\zeta^2}{\beta^2}) ) 
=\zeta p_{i,j}+ O(\frac{\zeta^2}{\beta})$.
Now, for $\beta$ sufficiently large,  we can always assume that the above error term  (i.e. the term  $  O(\frac{\zeta^2}{\beta})$)  can be made  smaller than $\frac{\epsilon\zeta}{2}$  and therefore  $\PP(y_{i,j}>\epsilon)<N^{\frac{\zeta\epsilon}{2}}$, with  $\zeta\frac{\epsilon}{2}>1$. 
This implies$\sum_i \PP(y_{i,j}>\epsilon) \le N^{1-\frac{\epsilon\zeta}{2}}\to 0$ as $N\to \infty$. As a consequence the  statement has been proved   for $\delta$ bounded away from 0, since, as result of previous  relationships we can choose $\beta= O(\frac{1}{\epsilon^2})$.   At last, for $\delta= o(1)$  by imposing that $ N^{1-\frac{\epsilon\zeta}{2}}>\delta$, 
we get that $ \beta(\epsilon, \delta)= O\left(\frac{1}{\epsilon^2}\frac{\log\frac{N}{\delta}}{\log N} \right)$ for the more general case. 

Similarly, the term $\sum_i\PP(-y_{i,j}>\epsilon)$ also tends to 0 as $N$ grows.
As for the second claim of the proposition we can write again
$\PP\left(\sup_{(i,j) \in \Ec}|z_{i,j}|>\epsilon'\right)\le \sum_i\PP(z_{i,j}>\epsilon)+\sum_i\PP(-z_{i,j}>\epsilon)$.
We then recall that $z_{i,j}=\widehat{d}_{i,j}-d_{i,j}$, $\widehat{d}_{i,j}=F^{-1}(y_{i,j}+p_{i,j})$, and $d_{i,j}=q_i-q_j$.
It follows that 
\begin{eqnarray}
  \PP(z_{i,j}>\epsilon') &=&   \PP\left(F^{-1}(y_{i,j}+p_{i,j}) -(q_i-q_j)>\epsilon'\right) \non
  &=&\PP\left(F^{-1}(y_{i,j}+F(q_i-q_j)) >\epsilon'+q_i-q_j\right)\non
    &=& \PP\left(y_{i,j}+F(q_i-q_j) >F(\epsilon'+q_i-q_j)\right)\non
    &=& \PP\left(y_{i,j} >F(\epsilon'+q_i-q_j)- F(q_i-q_j)\right) 
\end{eqnarray}
By defining $\epsilon\triangleq F(\epsilon'+F(q_i-q_j))- F(q_i-q_j)>0$ the convergence of $\PP(z_{i,j}>\epsilon')$ to 0 as $N$ grows immediately follows.
Similarly, it is straightforward to prove the convergence to 0 of the term $\sum_i\PP(-z_{i,j}>\epsilon)$.

\end{document}